\documentclass[structabstract]{aa}  
\usepackage{url,cancel,graphics,latexsym,epsfig,graphicx,amssymb,lscape,color,txfonts,natbib}
\usepackage[ruled,vlined,linesnumbered]{algorithm2e}     

\renewcommand{\d}[0]{{\rm d}}
\newcommand{\e}[0]{{\rm e}}
\renewcommand{\i}[0]{{\rm i}}
\newcommand{\ave}[1]{\langle #1 \rangle}
\newcommand{\Ave}[1]{\Big\langle #1 \Big\rangle}
\newcommand{\Ref}[1]{(\ref{#1})}

\newcommand{\mat}[1]{\tens{#1}}
\newcommand{\pprime}[0]{{\prime\prime}}

\newcommand{\glam}[0]{\mbox{GLAM}~}

\newcommand{\tr}[1]{{\rm tr}\left(#1\right)}

\newcommand*{\Scale}[2][4]{\scalebox{#1}{$#2$}}

\begin{document}

\title{Weak-lensing shear estimates with general adaptive moments, and studies 
of bias by pixellation, PSF distortions, and noise}

\author{Patrick Simon and Peter Schneider}

\institute{Argelander-Institut f\"ur Astronomie, Universit\"at Bonn, Auf dem
  H\"ugel 71, 53121 Bonn, Germany\\
  \email{psimon@astro.uni-bonn.de}}

\date{Received \today}

\authorrunning{P. Simon and P. Schneider} 

\titlerunning{Study of bias in lensing shape-measurements}

\def\LaTeX{L\kern-.36em\raise.3ex\hbox{a}\kern-.15em
    T\kern-.1667em\lower.7ex\hbox{E}\kern-.125emX}

  \abstract{%
    In weak gravitational lensing, weighted quadrupole moments of the
    brightness profile in galaxy images are a common way to estimate
    gravitational shear. We employ general adaptive moments (\glam\!)
    to study causes of shear bias on a fundamental level and for a
    practical definition of an image ellipticity. The \glam
    ellipticity has useful properties for any chosen weight profile:
    the weighted ellipticity is identical to that of isophotes of
    elliptical images, and in absence of noise and pixellation it is
    always an unbiased estimator of reduced shear.  We show that
    moment-based techniques, adaptive or unweighted, are similar to a
    model-based approach in the sense that they can be seen as
    imperfect fit of an elliptical profile to the image. Due to
    residuals in the fit, moment-based estimates of ellipticities are
    prone to underfitting bias when inferred from observed images. The
    estimation is fundamentally limited mainly by pixellation which
    destroys information on the original, pre-seeing image. We give an
    optimized estimator for the pre-seeing \glam ellipticity and
    quantify its bias for noise-free images. To deal with images where
    pixel noise is prominent, we consider a Bayesian approach to infer
    \glam ellipticity where, similar to the noise-free case, the
    ellipticity posterior can be inconsistent with the true
    ellipticity if we do not properly account for our ignorance about
    fit residuals. This underfitting bias, quantified in the paper,
    does not vary with the overall noise level but changes with the
    pre-seeing brightness profile and the correlation or heterogeneity
    of pixel noise over the image. Furthermore, when inferring a
    constant ellipticity or, more relevantly, constant shear from a
    source sample with a distribution of intrinsic properties (sizes,
    centroid positions, intrinsic shapes), an additional, now
    noise-dependent bias arises towards low signal-to-noise if
    incorrect prior densities for the intrinsic properties are
    used. We discuss the origin of this prior bias.  With regard to a
    fully-Bayesian lensing analysis, we point out that passing tests
    with source samples subject to constant shear may not be
    sufficient for an analysis of sources with varying shear.}


\maketitle


\section{Introduction}

Over the past decade measurements of distortions of galaxy images by
the gravitational lensing effect have developed into an important,
independent tool for cosmologists to study the large-scale
distribution of matter in the Universe and its expansion history
\citep[recent
reviews:][]{2006glsw.conf..269S,2008PhR...462...67M,2008ARNPS..58...99H,2010RPPh...73h6901M,2014arXiv1411.0115K}. These
studies exploit the magnification and shear of galaxy light bundles by
the tidal gravitational field of intervening matter. The shear gives
rise to a detectable coherent distortion pattern in the observed
galaxy shapes. The distortions are usually weak, only of order of a
few per cent of the unlensed shape of a typical galaxy
image. Therefore, the key to successfully devising gravitational shear
as cosmological tool are accurate measurements of the shapes of many,
mostly faint and hardly resolved galaxy images.

There has been a boost of interest in methods of shape measurements in
anticipation of the gravitational lensing analysis of the upcoming
next generation of wide-field imaging surveys \citep[e.g.,
Euclid;][]{2011arXiv1110.3193L}. Despite being sufficient for
contemporary surveys, current methodologies are not quite at the
required level of accuracy to fully do justice to the amount of
cosmological information in future lensing surveys
\citep{2006MNRAS.368.1323H,2007MNRAS.376...13M}.  The challenge all
methodologies face is that observable, noisy (post-seeing) galaxy
images are modifications of the actual (pre-seeing) image owing to
instrumental and possible atmospheric effects. Post-seeing galaxy
images are subject to pixellation as well as instrumental noise, sky
noise, photon noise, and random overlapping with very faint objects
\citep{2012MNRAS.423.3163K,2015MNRAS.449..685H}. In addition, galaxies
are not intrinsically circular such that their ellipticities are noisy
estimators of the cosmic distortion.  Current theoretical work
consequently focuses on sources of bias in shape measurements, such as
pixel-noise bias, shape-noise bias, underfitting bias, colour
gradients, or several selection biases
\citep{2003MNRAS.343..459H,2004MNRAS.353..529H,2005MNRAS.361.1287M,2010A&A...510A..75M,2011MNRAS.410.2156V,2012MNRAS.424.2757M,2012MNRAS.427.2711K,2013MNRAS.429..661M,2013MNRAS.432.2385S}.

One major source of bias is the pixel-noise bias or simply noise bias
hereafter. This bias can at least partly be blamed on the usage of
point estimates of galaxy shapes in a statistical analysis, i.e.,
single-value estimators of galaxy ellipticities
\citep{2012MNRAS.425.1951R}. This begs the question whether it is
feasible to eradicate noise bias by means of a more careful treatment
of the statistical uncertainties in the measurement of galaxy
ellipticities within a fully Bayesian framework. Indeed recent
advances in image processing for weak gravitational lensing strongly
support this idea, at least for the inference of constant shear
(\citealt{2014MNRAS.444L..25S}; \citealt{2014MNRAS.438.1880B}, BA14
hereafter; \citealt{2015arXiv150805655B}, BAKM16 hereafter).  In
contrast, the contemporary philosophy with point estimates is to
perform elaborate, time-consuming calibrations of biased estimators by
means of simulated images; the calibration accuracy is, additionally,
only as good as the realism of simulated images
\citep[e.g.,][]{2015MNRAS.449..685H}. To be fair, code implementations
of non-Bayesian techniques are typically an order of magnitude or more
faster than Bayesian codes which could be a decisive factor for
upcoming surveys.

We take here a new look into possible causes of bias in shear
measurements on a fundamental level.  To this end, we examine,
step-by-step with increasing complexity, a fully-Bayesian lensing
analysis based on weighted brightness moments of galaxy images
\citep{gelman2003bayesian,2003Book...MACKAY}. While the method in BA14
and BAKM16 is set in Fourier space, we work with moments in angular
space which has benefits in the case of correlated noise or missing
pixels in realistic images.  Moment-based methods as ours are
non-parametric; this means they are free from assumptions about the
galaxy brightness profile. They hence appear to be advantageous for
reducing bias, but nonetheless the specific choice of the adaptive
weight for the moments is known to produce bias
\citep{2014MNRAS.439.1909V,2010MNRAS.404..458V}. The origin of this
problem, which principally also affects unweighted moments, becomes
obvious in our formalism.  We define as practical measure of galaxy
shape a generalization of the impractical ellipticity $\epsilon$
expressed in terms of unweighted moments \citep[][SS97
hereafter]{1995ApJ...449..460K,1997A&A...318..687S}.  Being Bayesian,
our measurement of ellipticity results in a Monte-Carlo sample of the
probability distribution function (PDF) of $\epsilon$ which should be
propagated in a fully Bayesian analysis.  That is: we do not devise
point estimators in order to ideally stay clear of noise bias. This
overall approach of general adaptive moments, \glam hereafter, is
inspired by and comparable to \citet{2002AJ....123..583B} apart from
the Bayesian framework and some technical differences: (i) for any
adaptive weight, the perfectly measured \glam ellipticity is an
unbiased estimator of gravitational shear unaffected by shape-noise
bias; (ii) the adaptive weight may have a non-Gaussian radial profile;
(iii) our inference of the pre-seeing ellipticity is realised as
forward-fitting of elliptical profiles (so-called templates), that is
we do not determine the brightness moments of the post-seeing image
and correct them to estimate the pre-seeing moments
\citep[cf.][]{2003MNRAS.343..459H,2005MNRAS.361.1287M}.

As a disclaimer, the \glam methodology outlined here is prone to bias,
even where a fully Bayesian analysis can be realised, and, at this
stage, its performance is behind that of other techniques. The aim of
this paper is to elucidate causes of bias, instead of proposing a new
technique that is competitive to existing techniques. However, these
findings are also relevant for other methodologies because model-based
or moment-based approaches are linked to \glam\!\!.

For this paper, we exclude bias from practically relevant factors: the
insufficient knowledge of the PSF or noise properties, blending of
images, and the selection of source galaxies \citep[see
e.g.,][]{2006MNRAS.368.1323H,2011A&A...528A..51H,2014arXiv1406.1506D}. We
focus on the core of the problem of shape measurements which is the
inference of pre-seeing ellipticities from images whose full
information on the brightness profile have been lost by instrumental
limitations.
  
The paper is laid out as follows. In Sect. 2, we introduce the
formalism of \glam for a practical definition of ellipticity with
convenient transformation properties under the action of gravitational
shear.  We also analytically investigate the limits of measuring the
pre-seeing ellipticity from a noise-free but both PSF-convolved and
pixellated image.  In Sect. 3, we construct a statistical model for
the \glam ellipticity of noisy post-seeing images.  We then study the
impact of inconsistencies in the posterior model of ellipticity in
three, increasingly complex scenarios. First, we analyse with
independent exposures of the same pre-seeing image the ellipticity
bias due to a misspecified likelihood in the posterior (underfitting
bias). Second, we consider samples of noisy images with the same
ellipticity but distributions of intrinsic properties such as sizes or
centroid positions. Here a new contribution to the ellipticity bias
emerges if the prior densities of intrinsic properties are incorrectly
specified (prior bias). Third in Sect. 4, we perform numerical
experiments with samples of noisy galaxy images of random intrinsic
shapes that are subject to constant shear. With these samples we study
the impact of inconsistent ellipticity posteriors on shear constraints
(shear bias). We also outline details on our technique to Monte-Carlo
sample ellipticity or shear posteriors.  We discuss our results and
possible improvements of the \glam approach in Sect. 5.


\section{General adaptive moments}

\subsection{Definition}

Let $I(\vec{x})$ be the light distribution in a galaxy image of
infinite resolution and without noise where $\vec{x}$ is the position
on the sky. A common way of defining a (complex) ellipticity of a
galaxy image uses the quadrupole moments
\begin{equation}
  \label{eq:unwqij}
 Q_{ij}=
 \frac{\int\d^2x\;(x_i-X_{0,i})\,(x_j-X_{0,j})\,I(\vec{x})}
 {\int\d^2x\;I(\vec{x})}
\end{equation}
of $I(\vec{x})$ relative to a centroid position 
\begin{equation}
  \label{eq:uwx0}
  \vec{X}_0=
  \frac{\int\d^2x\;\vec{x}\,I(\vec{x})}
  {\int\d^2x\;I(\vec{x})}
\end{equation}
of the image \citep[e.g.,][]{2001PhR...340..291B}. For real
applications, the quadrupole moments are soundly defined only if they
involve a weight, decreasing with separation from the centroid
position $\vec{X}_0$, because galaxies are not isolated so that the
normalisation $\int\d^2x\;I(\vec{x})$ and the brightness moments
diverge. \citet[][H03 hereafter]{2003MNRAS.343..459H} address the
divergence problems by realising an adaptive weighting scheme by
minimising the error functional, sometimes dubbed the energy
functional,
\begin{equation}
   \label{eq:functional}
   E(\vec{p}|I)=
   \frac{1}{2}
   \int\d^2x\;
   \Big[I(\vec{x})-Af(\rho)\Big]^2\;,
\end{equation}
with the quadratic form
\begin{equation}
  \rho:=
  (\vec{x}-\vec{x}_0)^{\rm T}\mat{M}^{-1}(\vec{x}-\vec{x}_0)
\end{equation}
and the second-order tensor
\begin{equation}
  \label{eq:ellipticity}
  \mat{M}=\frac{T}{2}
  \left(
    \begin{array}{cc}
      1+e_1 & e_2 \\
      e_2 & 1-e_1
    \end{array}
  \right)\;.
\end{equation}
The tensor $\mat{M}$ is expressed in terms of the complex ellipticity
\mbox{$e=e_1+\i e_2$} and the size $T$ of the image; $f(\rho)$ is a
weight function that HS03 chose to be a Gaussian weight
\mbox{$f(\rho)=\e^{-\rho/2}$}.  In comparison to HS03, we have
slightly changed the definition of $\rho$ for convenience: here we use
$\rho$ instead of $\rho^2$. The set $\vec{p}=(A,\vec{x}_0,\mat{M})$,
comprises a set of six parameters on which the functional $E$ depends
for a given galaxy image $I$.

Frequently another definition of complex ellipticity, the
$\epsilon$-ellipticity, is more convenient (sometimes also known as
the third flattening). It arises if we write $\rho$ in the form
\begin{equation}
  \label{eq:rho}
  \rho
  =\left|\mat{V}^{-1}(\vec{x}-\vec{x}_0)\right|^2
  =(\vec{x}-\vec{x}_0)^{\rm T}\mat{V}^{-2}(\vec{x}-\vec{x}_0)\;,
\end{equation}
where $\mat{V}$ is symmetric. Obviously, we have $\mat{V}^2=\mat{M}$
or $\mat{V}=\sqrt{\mat{M}}$. By writing $\mat{V}$ in the form
\begin{equation}
  \mat{V}=
  \frac{t}{2}
  \left(
    \begin{array}{cc}
      1+\epsilon_1 & \epsilon_2\\
      \epsilon_2 & 1-\epsilon_1
    \end{array}
  \right)
\end{equation}
we see that $\mat{V}^2=\mat{M}$ implies $2T=t^2\,(1+|\epsilon|^2)$,
and
\begin{equation}
  \label{eq:elltransf}
  e
  =\frac{2\epsilon}{1+|\epsilon|^2}~;~
  \epsilon=
  \epsilon_1+\i\epsilon_2=\frac{e}{1+\sqrt{1-|e|^2}}\;.
\end{equation}
We henceforth use $\epsilon$ as parametrisation of $\mat{M}$ because
$\epsilon$ is an unbiased estimator of reduced shear in the absence of
a PSF and pixellation (SS97). Conversely, the ellipticity $e$ has to
be calibrated with the distribution of unsheared ellipticities which
poses another possible source of bias in a lensing analysis (H03).

As generally derived in Appendix \ref{sect:gam}, the parameters
$\vec{p}$ at the minimum of \Ref{eq:functional} are:
\begin{equation}
  \label{eq:amx0}
  \vec{x}_0 =
  \frac{\int\d^2x\;\vec{x}\,I(\vec{x})f^\prime(\rho)}
  {\int\d^2x\;I(\vec{x})\,f^\prime(\rho)}~;~
  A=
  \frac{\int\d^2x\;I(\vec{x})f(\rho)}
  {\int\d^2x\;f^2(\rho)}\
\end{equation}
and 
\begin{equation}
  \label{eq:amm}
  \mat{M} =
  \beta
  \,\frac{\int\d^2x\;
    (\vec{x}-\vec{x}_0)(\vec{x}-\vec{x}_0)^{\rm T}
    I(\vec{x})f^\prime(\rho)}
  {\int\d^2x\;I(\vec{x})f(\rho)}\;,
\end{equation}
where $\beta:=-2\int_0^\infty\d s\;f^2(s)/f^2(0)$ is a constant.
These equations are derived by HS03 for a Gaussian $f(\rho)$, for
which we have \mbox{$f^\prime(\rho):=\d f/\d\rho=-f/2$}. This shows
that the best-fitting $f(\rho)$ has the same centroid and, up to a
scalar factor, the same second moment $\mat{M}=\mat{V}^2$ as the with
$f^\prime(\rho)$ adaptively weighted image $I(\vec{x})$. This is
basically also noted in \cite{2009MNRAS.398..471L} where it is argued
that the least-square fit of \emph{any} sheared model to a pre-seeing
image $I(\vec{x})$ provides an unbiased estimate of the shear, even if
it fits poorly.

In this system of equations, the centroid $\vec{x}_0$ and tensor
$\mat{M}$ are implicitly defined because both $f(\rho)$ and
$f^\prime(\rho)$ on the right-hand-side are functions of the unknowns
$\vec{x}_0$ and $\mat{M}$: the weights \emph{adapt} to the position,
size, and shape of the image. The Eqs. \Ref{eq:amx0} and \Ref{eq:amm}
therefore need to be solved iteratively. The iteration should be
started at a point which is close to the final solution. Such a
starting point could be obtained by using the image position,
determined by the image detection software, as initial value for
$\vec{x}_0$, and tensor $\mat{M}$ determined from a circular weight
function with the same functional form as $f$. Nevertheless, there is
no guarantee that the solution of this set of equations is unique. In
fact, for images with two brightness peaks one might suspect that
there are multiple local minima in $\vec{p}$ of the functional
$E$. This may occur, for instance, in the case of blended images. One
standard solution to this particular problem is to identify blends and
to remove these images from the shear catalogue. Alternatively we
could in principle try to fit two template profiles to the observed
image, i.e., by adding a second profile $E_2(\vec{p}_2|I)$ to the
functional \Ref{eq:functional} and by minimising the new functional
with respect to the parameter sets $\vec{p}$ and $\vec{p}_2$ of both
profiles simultaneously.

\subsection{Interpretation}
\label{sect:interpretation}

If an image $I(\vec{x})$ has confocal elliptical isophotes, with the
same ellipticity for all isophotes, one can define the ellipticity of
the image uniquely by the ellipticity of the isophotes. In this case,
the ellipticity $\epsilon$ defined by the minimum of
\Ref{eq:functional} coincides with the ellipticity of the isophotes
for any weight $f$. We show this property in the following.

Assume that the brightness profile $I(\vec{x})$ is constant on
confocal ellipses so that we can write $I(\vec{x})=S(\zeta)$ where
$\zeta:=(\vec{x}-\vec{x}_{\rm c})^{\rm
  T}\mat{B}^{-2}(\vec{x}-\vec{x}_{\rm c})$.
Here $\vec{x}_{\rm c}$ denotes the centre of the image, and the matrix
elements of $\mat{B}$ describe the size and the shape of the image, in
the same way as we discussed for the matrix $\mat{V}$ before. The
function $S(\zeta)$ describes the radial brightness profile of the
image. We start by writing \Ref{eq:amx0} in the form
\begin{equation}
  0=\int\d^2x\;S(\zeta)\,f^\prime(\rho)\,(\vec{x}-\vec{x}_0)\;,
\end{equation}
and introduce the transformed position vector
$\vec{z}=\mat{B}^{-1}(\vec{x}-\vec{x}_{\rm c})$, or
$\vec{x}=\mat{B}\vec{z}+\vec{x}_{\rm c}$. Then the previous equation
becomes
\begin{equation}
  \label{eq:centroid1}
  0=\int\d^2z\;S(|\vec{z}|^2)\,f^\prime(\rho)\,(\mat{B}\vec{z}+\vec{x}_{\rm
    c}-\vec{x}_0)\;,
\end{equation}
where in terms of $\vec{z}$ the quadratic form $\rho$ is
\begin{equation}
  \rho=(\mat{B}\vec{z}+\vec{x}_{\rm c}-\vec{x}_0)^{\rm
    T}\mat{V}^{-2}(\mat{B}\vec{z}+\vec{x}_{\rm c}-\vec{x}_0)\;.
\end{equation}
From these equations, we can see that $\vec{x}_0=\vec{x}_{\rm c}$ is
the solution of \Ref{eq:centroid1} since then $\rho$ is an even
function of $\vec{z}$, $S$ is an even function of $\vec{z}$, whereas
the term in the parenthesis of \Ref{eq:centroid1} is odd, and the
integral vanishes due to symmetry reasons. Thus we found that our
adaptive moments approach yields the correct centre of the image.

Next, we rewrite \Ref{eq:amm} in the form
\begin{eqnarray}
  \nonumber
  \mat{V}^2\int\d^2x\;I(\vec{x})f(\rho)&=&
  \beta\int\d^2x\;(\vec{x}-\vec{x}_0)\,(\vec{x}-\vec{x}_0)^{\rm T}
  I(\vec{x})\,f^\prime(\rho)\\
  &=&  \label{eq:secondorder1}
  \beta\det{\mat{B}}\int\d^2z\;S(|\vec{z}|^2)\,f^\prime(\rho)\,\mat{B}\vec{z}\vec{z}^{\rm
    T}\mat{B}\;,
\end{eqnarray}
where $\beta$, see Eq. \Ref{eq:amm}, is a constant factor (see
Appendix \ref{ap:beta}). We again used the transformation from
$\vec{x}$ to $\vec{z}=\mat{B}^{-1}(\vec{x}-\vec{x}_{\rm c})$ and
employed the fact that $\vec{x}_0=\vec{x}_{\rm c}$. Accordingly, we
have $\rho=\vec{z}^{\rm T}\mat{B}\mat{V}^{-2}\mat{B}\vec{z}$. We now
show that the solution of \Ref{eq:secondorder1} is given by
$\mat{V}=\lambda\mat{B}$ with $\lambda$ being a scalar factor. Using
this Ansatz, we get $\rho=\lambda^{-2}|\vec{z}|^2$, and
\Ref{eq:secondorder1} can be written, after multiplying from the left
and from the right by $\mat{B}^{-1}$, as
\begin{equation}
  \label{eq:secondorder2}
  \mat{B}^{-1}\mat{V}^2\mat{B}^{-1}=\lambda^2\mat{1}
  =
  \beta\frac{\int\d^2z\;S(|\vec{z}|^2)f^\prime(|\vec{z}|^2/\lambda^2)\vec{z}\vec{z}^{\rm
      T}}
  {\int\d^2z\;S(|\vec{z}|^2)f(|\vec{z}|^2/\lambda^2)}\;.
\end{equation}
Since both $S$ and $f^\prime$ in the numerator depend solely on
$|\vec{z}|^2$, the tensor on the right hand side is proportional to
the unit tensor $\mat{1}$, and \Ref{eq:secondorder2} becomes a scalar
equation for the scalar $\lambda$,
\begin{equation}
  \lambda^2=
  \beta\frac{\int\d s\;s\,S(s)f^\prime(s/\lambda^2)}
  {\int\d s\;S(s)f(s/\lambda^2)}\;,
\end{equation}
whose solution depends on the brightness profile $S$ and the chosen
weight function $f$. However, the fact that $\mat{B}$ differs from
$\mat{V}$ only by the scalar factor $\lambda$ implies that the derived
ellipticity $\epsilon$ of $\mat{V}$ is the same as that of the
elliptical image. Therefore we have shown that the approach of
adapted moments recovers the true ellipticity with elliptical
isophotes for \emph{any} radial weight function $f$.

For a general brightness profile of the image, the interpretation of
the \glam ellipticity $\epsilon$ is less clear, and the ellipticity
generally depends on the weight $f$. Nevertheless, $\epsilon$ is
uniquely defined as long as a minimum of the functional
\Ref{eq:functional} can be found. More importantly, for \emph{any}
weight $f$ the \glam ellipticity obeys the same simple transformation
law under the action of gravitational shear, as shown in the following
section.

\subsection{Transformation under shear}
\label{sect:shear}

We now consider the effect of a shear \mbox{$\gamma=\gamma_1+\i
  \gamma_2$} and convergence $\kappa$ on the \glam ellipticity
$\epsilon$ \citep{2001PhR...340..291B}. For an image with no noise, no
pixellation, and no PSF convolution the ellipticity $\epsilon$ should
be an unbiased estimate of the reduced shear \mbox{$g=g_1+\i
  g_2=\gamma\,(1-\kappa)^{-1}$}. This is clearly true for sources that
intrinsically have circular isophotes since the isophotes of the
sheared images are confocal ellipses with an ellipticity
\mbox{$\epsilon=g$}. The minimum of \Ref{eq:functional} is hence at
$g$ by means of the preceding discussion.

For general images, let $\epsilon_{\rm s}=\epsilon_{{\rm s},1}+\i
\epsilon_{{\rm s},2}$ be the complex ellipticity of the image in the
source plane and $\epsilon=\epsilon_1+\i \epsilon_2$ its complex
ellipticity in the lens plane.  We show now that for any brightness
profile and template $f(\rho)$, \glam ellipticities have the extremely
useful property to transform under the action of a reduced shear
according to
\begin{equation}
  \label{eq:ss95}
  \epsilon(g,\epsilon_{\rm s})=
  \left\{
    \begin{array}{ll}
      {\Scale[1.4]{\frac{\epsilon_{\rm s}+g}{1+g^\ast\epsilon_{\rm s}}}}
      & \;,\,|g|\le1\;,\\\\
      {\Scale[1.4]{\frac{1+\epsilon_{\rm s}^\ast\,g}{\epsilon_{\rm s}^\ast+g^\ast}}}
      & \;,\,|g|>1
    \end{array}
  \right.\;.
\end{equation}
This is exactly the well-known transformation obtained from unweighted
moments (SS97). To show this generally for \glam$\!\!$, let $I_{\rm
  s}(\vec{y})$ be the surface brightness of an image in the source
plane with source plane coordinates $\vec{y}$. The centroid
$\vec{y}_0$ and moment tensor $\mat{M}_{\rm s}$ of $I_{\rm s}$ are
defined by the minimum of \Ref{eq:functional} or, alternatively, by
the analog of Eqs. \Ref{eq:amx0} and \Ref{eq:amm} through
\begin{equation}
  \label{eq:y0center}
  \int\d^2y\;I_{\rm s}(\vec{y})f(\rho_{\rm s})(\vec{y}-\vec{y}_0)=0
\end{equation}
and 
\begin{equation}
  \label{eq:Ms}
  \mat{M}_{\rm s}=
  \beta\,\frac{\int\d^2y\;(\vec{y}-\vec{y}_0)(\vec{y}-\vec{y}_0)^{\rm
      T}I_{\rm s}(\vec{y})f^\prime(\rho_{\rm s})}
  {\int\d^2y\;I_{\rm s}(\vec{y})f(\rho_{\rm s})}
\end{equation}
with
\begin{equation}
  \label{eq:rhos}
  \rho_{\rm s}=(\vec{y}-\vec{y}_0)^{\rm T}\mat{M}_{\rm
    s}^{-1}(\vec{y}-\vec{y}_0)\;.
\end{equation}
The ellipticity $\epsilon_{\rm s}$ is given by
\begin{equation}
  \label{eq:ms1}
  \mat{M}_{\rm s}=\mat{V}_{\rm s}^2~;~
  \mat{V}_{\rm s}=
  \frac{t_{\rm s}}{2}
  \left(
    \begin{array}{cc}
      1+\epsilon_{{\rm s},1} & \epsilon_{{\rm s},2}\\
      \epsilon_{{\rm s},2} & 1-\epsilon_{{\rm s},1}
    \end{array}
  \right)\;.
\end{equation}

We now shear the image $I_{\rm s}$. The shear and the convergence are
assumed to be constant over the extent of the image, i.e., lens plane
positions $\vec{x}$ are linearly mapped onto source plane positions
$\vec{y}$ by virtue of $\vec{y}-\vec{y}_{\rm c}={\cal
  A}(\vec{x}-\vec{x}_{\rm c})$ where
\begin{equation}
  {\cal A}={\cal A}^{\rm T}=
  (1-\kappa)
  \left(
  \begin{array}{cc}
    1-g_1 & -g_2 \\
    -g_2 & 1+g_1
  \end{array}
  \right)\;;
\end{equation}
$\vec{x}_{\rm c}$ and $\vec{y}_{\rm c}$ are such that the point
$\vec{x}_{\rm c}$ is mapped onto the point $\vec{y}_{\rm c}$ by the
lens equation, and both are chosen to be the central points around
which the lens equation is linearised. We then find for the centroid
$\vec{y}_0$ in the source plane
\begin{equation}
  \vec{y}-\vec{y}_0=\vec{y}-\vec{y}_{\rm c}+\vec{y}_{\rm
    c}-\vec{y}_0
  ={\cal A}(\vec{x}-\vec{x}_1)\;,
\end{equation}
where $\vec{x}_1-\vec{x}_{\rm c}={\cal A}^{-1}(\vec{y}_0-\vec{y}_{\rm
  c})$. This then yields for \Ref{eq:rhos}
\begin{equation}
  \rho_{\rm s}=(\vec{x}-\vec{x}_1)^{\rm T}{\cal A}\mat{M}_{\rm
    s}^{-1}{\cal A}\,(\vec{x}-\vec{x}_1)\;.
\end{equation}
In the next step, the expression \Ref{eq:y0center} for the source
centre can be rewritten by transforming to the image coordinates and
using the conservation of surface brightness $I_{\rm
  s}(\vec{y}(\vec{x}))=I(\vec{x})$:
\begin{eqnarray}
  \label{eq:xcentre1}
  0&=&\int\d^2y\;I_{\rm s}(\vec{y})\,f(\rho_{\rm
    s})\,(\vec{y}-\vec{y}_0)\\
  \nonumber&=&
  \det{(\cal A)}\,{\cal
    A}\int\d^2x\;I(\vec{x})\,f(\rho_{\rm s})\,(\vec{x}-\vec{x}_1)\;.
\end{eqnarray}
With the same transformation, we rewrite the moment tensor \Ref{eq:Ms}
as
\begin{equation}
  \label{eq:xtensor1}
  \mat{M}_{\rm s}=
  \beta{\cal A}\,
  \frac{\int\d^2x\;(\vec{x}-\vec{x}_1)(\vec{x}-\vec{x}_1)^{\rm
      T}I(\vec{x})f^\prime(\rho_{\rm s})}
  {\int\d^2x\;I(\vec{x})f(\rho_{\rm s})}\,
  {\cal A}\;.
\end{equation}
On the other hand, minimising the functional \Ref{eq:functional} for
the surface brightness $I(\vec{x})$ in the lens plane yields the
expressions \Ref{eq:amx0} and \Ref{eq:amm} for $\vec{x}_0$ and
$\mat{M}$, respectively. We then see that the Eqs. \Ref{eq:xcentre1},
\Ref{eq:xtensor1} and \Ref{eq:amx0}, \Ref{eq:amm} agree with each
other, if we set
\begin{equation}
  \mat{M}_{\rm s}={\cal A}\mat{M}{\cal A}~;~
  \vec{x}_1=\vec{x}_0\;,
\end{equation}
for which $\rho_{\rm s}=\rho$. In particular, this shows that the
centre $\vec{x}_0$ of the image is mapped onto the centre $\vec{y}_0$
of the source.

The relation
\begin{equation}
  \mat{M}={\cal A}^{-1}\mat{M}_{\rm s}{\cal A}^{-1}
\end{equation}
can be rewritten in terms of the square roots of the matrix $\mat{M}$
as
\begin{equation}
  \label{eq:MsVs}
  \mat{M}_{\rm s}\equiv\mat{V}^2_{\rm s}={\cal A}\mat{V}^2{\cal
    A}
\end{equation}
where \mbox{$\mat{V}_{\rm s}=\sqrt{\mat{M}_{\rm s}}$} is uniquely
defined by requiring that for the symmetric, positive-definite matrix
$\mat{M}_{\rm s}$, $\mat{V}_{\rm s}$ is symmetric and
positive-definite. Although both $\mat{V}$ and $\cal A$ are symmetric,
$\mat{V}{\cal A}$ is in general not. Therefore $\mat{V}_{\rm s}$
cannot be readily read off from \Ref{eq:MsVs}. Instead, we use a
rotation matrix
\begin{equation}
  \mat{R}(\varphi)=
  \left(
    \begin{array}{cc}
      \cos{\varphi} & -\sin{\varphi} \\
      +\sin{\varphi} & \cos{\varphi}
    \end{array}
  \right)~;~
  \mat{R}^{\rm T}(\varphi)=\mat{R}^{-1}(\varphi)=\mat{R}(-\varphi)\;,
\end{equation}
to write
\begin{equation}
  \mat{V}_{\rm s}^2
  =
  {\cal A}\mat{V}\mat{R}^{-1}(\varphi)\mat{R}(\varphi)\mat{V}{\cal A}
  =
  [\mat{R}(\varphi)\mat{V}{\cal A}]^{\rm T}[\mat{R}(\varphi)\mat{V}{\cal A}]\;.
\end{equation}
If we now choose $\varphi$ to be such that
$\mat{R}(\varphi)\mat{V}{\cal A}$ is symmetric, then $\mat{V}_{\rm
  s}=\mat{R}(\varphi)\mat{V}{\cal A}$. After a bit of algebra, we find
that the rotation angle $\varphi$ is given through
\begin{equation}
  \e^{-\i\varphi}=
  \frac{1-\epsilon g^\ast}{|1-\epsilon g^\ast|}\;,
\end{equation}
and we obtain as final result $\mat{V}_{\rm s}$ as in \Ref{eq:ms1}
with
\begin{equation}
  \label{eq:epstrans}
  t_{\rm s}=|1-\epsilon g^\ast|\,t~;~
  \epsilon^{\rm
    s}=\frac{\epsilon-g}{1-\epsilon g^\ast}\;.
\end{equation}
The inverse of \Ref{eq:epstrans} is given by
\begin{equation}
  \label{eq:einverse}
  \epsilon=
  \frac{\epsilon_{\rm s}+g}{1+\epsilon_{\rm
      s}g^\ast}\;.
\end{equation}
We recover for the \glam ellipticity $\epsilon$ exactly the
transformation law of unweighted moments (SS97). The \glam ellipticity
$\epsilon$ is therefore an unbiased estimator of the reduced shear $g$
along the line-of-sight of the galaxy, and there is no need to
determine unweighted moments.

As a side remark, the transformation between $\epsilon$ and
$\epsilon_{\rm s}$ is a linear conformal mapping from the unit circle
onto the unit circle, and from the origin of the $\epsilon$-plane onto
the point $-g$ in the $\epsilon_{\rm s}$-plane. If \mbox{$|g|>1$},
then $g$ has to be replaced by $1/g^\ast$ in Eq. \Ref{eq:einverse},
but we shall not be concerned here with this situation in the strong
lensing regime.

\subsection{Point spread function and pixellation}
\label{sect:psfpixel}

We have defined the \glam ellipticity $\epsilon$ of an image
$I(\vec{x})$ relative to an adaptive weight $f(\rho)$. This definition
is idealized in the sense that it assumes an infinite angular
resolution and the absence of any atmospheric or instrumental
distortion of the image. Equally important, it ignores pixel noise. In
this section, we move one step further to discuss the recovery of the
original $\epsilon$ of an image after it has been convolved with a PSF
and pixellated. The problem of properly dealing with noise in the
image is discussed subsequently.

Let $I_{\rm pre}(\vec{x})$ be the original image prior to a PSF
convolution and pixellation. This we call the `pre-seeing'
image. Likewise, by the vector $\vec{I}_{\rm post}$ of $N_{\rm pix}$
values we denote the `post-seeing' image that has been subject to a
convolution with a PSF and pixellation.  For mathematical convenience,
we further assume that $I_{\rm pre}(\vec{x})$ is binned on a fine
auxiliary grid with \mbox{$N\gg N_{\rm pix}$} pixels of solid angle
$\Omega$. We list these pixel values as vector $\vec{I}_{\rm
  pre}$. The approximation of $I_{\rm pre}(\vec{x})$ by the vector
$\vec{I}_{\rm pre}$ becomes arbitrarily accurate for
\mbox{$N\to\infty$}. Therefore we express the post-seeing image
\mbox{$\vec{I}_{\rm post}=\mat{L}\vec{I}_{\rm pre}$} by the linear
transformation matrix $\mat{L}$ applied to the pre-seeing image
$\vec{I}_{\rm pre}$. The matrix $\mat{L}$ with $N\times N_{\rm pix}$
elements combines the effect of a (linear) PSF convolution and
pixellation. Similarly, we bin the template $f(\rho)$ in the
pre-seeing frame to the grid of $\vec{I}_{\rm pre}$, and we denote the
binned template by the vector $\vec{f}_\rho$; as usual, the quadratic
form $\rho$ is here a function of the variables
$(\vec{x}_0,\epsilon,t)$, Eq. \Ref{eq:rho}. The \glam parameters
$\vec{p}_{\rm pre}$ of the pre-seeing image are given by the minimum
of $E(\vec{p}|I_{\rm pre})$, or approximately by
\begin{equation}
  \label{eq:highresE}
  E_{\rm pre}(\vec{p}|\vec{I}_{\rm pre}):=
  (\vec{I}_{\rm pre}-A\vec{f}_\rho)^{\rm T}(\vec{I}_{\rm pre}-A\vec{f}_\rho)
  \approx 2\Omega^{-1}E(\vec{p}|I_{\rm pre})
  \;.
\end{equation}
For the recovery of the pre-seeing ellipticity $\epsilon$, the
practical challenge is to derive the pre-seeing parameters
$\vec{p}_{\rm pre}$ from the observed image $\vec{I}_{\rm post}$ in
the post-seeing frame. For this task, we assume that the
transformation $\mat{L}$ is exactly known.  Note that a linear mapping
$\mat{L}$ is an approximation here; we ignore the nonlinear effects in
the detector
\citep{2014JInst...9C4001P,2015arXiv150102802G,2015ExA...tmp...15N,2015MNRAS.449.2219M}.

For a start, imagine a trivial case where no information is lost by
going from $\vec{I}_{\rm pre}$ to $\vec{I}_{\rm post}$. We express
this case by a transformation $\mat{L}$ that can be inverted, i.e., we
have \mbox{$N=N_{\rm pix}$} and $\mat{L}$ is regular. We then obtain
$\vec{p}_{\rm pre}$ by minimising
$E_{\rm pre}(\vec{p}|\mat{L}^{-1}\vec{I}_{\rm post})$ with respect to
$\vec{p}$: we map $\vec{I}_{\rm post}$ to the pre-seeing frame and
analyse $\vec{I}_{\rm pre}=\mat{L}^{-1}\vec{I}_{\rm post}$ there. This
is equivalent to minimising the form
\mbox{$(\vec{I}_{\rm post}-A\mat{L}\vec{f}_\rho)^{\rm
    T}(\mat{L}\mat{L}^{\rm T})^{-1}(\vec{I}_{\rm
    post}-A\mat{L}\vec{f}_\rho)$} in the post-seeing frame.

\begin{figure}
  \begin{center}
    \epsfig{file=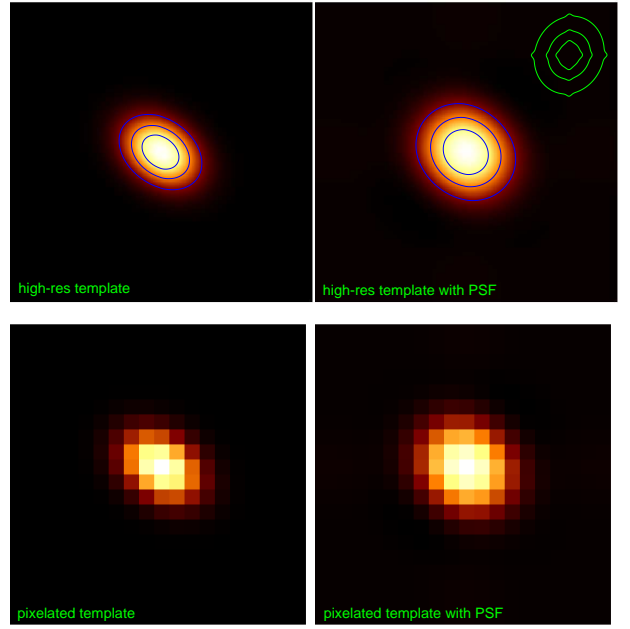,width=80mm,angle=0}
  \end{center}
  \caption{\label{fig:templexamp} Examples of \glam templates in the
    pre-seeing frame, $\vec{f}_\rho$ (top left), and the post-seeing
    frame, $\mat{L}\vec{f}_\rho$ (other panels); the templates are
    Gaussian radial profiles with $f(\rho)=\e^{-\rho/2}$. The bottom
    left panel simulates only pixellation, whereas the right column
    also shows the impact of a PSF, indicated in the top right corner,
    without (top) and with pixellation (bottom).}
\end{figure}

For realistic problems where $\mat{L}^{-1}$ does not exist, because
$N\gg N_{\rm pix}$, this trivial case at least suggests to determine
the minimum $\vec{p}_{\rm post}$ of the new functional
\begin{equation}
  \label{eq:functional2}
  E_{\rm post}(\vec{p}|\vec{I}_{\rm post}):=
  (\vec{I}_{\rm post}-A\mat{L}\vec{f}_\rho)^{\rm T}\mat{U}
  (\vec{I}_{\rm post}-A\mat{L}\vec{f}_\rho)
\end{equation}
as estimator of $\vec{p}_{\rm pre}$. This way we are setting up an
estimator by forward-fitting the template $\vec{f}_\rho$ to the image
in the post-seeing frame with the matrix $\mat{U}$ being a metric for
the goodness of the fit. Clearly, should $\mat{L}^{-1}$ exist we
recover \Ref{eq:highresE} only by adopting
$\mat{U}=(\mat{L}\mat{L}^{\rm T})^{-1}$. So we could equivalently
obtain $\vec{p}_{\rm pre}$, without bias, by fitting
$\mat{L}\vec{f}_{\rho}$ to the observed image $\vec{I}_{\rm post}$ in
this case. However, realistically $\mat{L}$ is singular: the recovery
of $\vec{p}_{\rm pre}$ from \Ref{eq:functional2} can only be done
approximately.  Then we could at least find an optimal metric to
minimise the bias. We return to this point shortly. In any case, the
metric has to be positive-definite and symmetric such that always
\mbox{$E_{\rm post}\ge0$}. Note that the moments at the minimum of
\Ref{eq:functional2} are related but not identical to the adaptive
moments in the post-seeing frame. To obtain the latter we would fit a
pixellated $\vec{f}_\rho$ with $\mat{U}=\mat{1}$ to $\vec{I}_{\rm
  post}$. The bottom and top right images in Fig. \ref{fig:templexamp}
display examples of post-seeing templates that are fitted to a
post-seeing image to estimate $\vec{p}_{\rm pre}$ with the functional
\Ref{eq:functional2}.

For singular $\mat{L}$, the minimum of the functional yields an
unbiased $\vec{p}_{\rm pre}$ for any $\mat{U}$ if
\begin{enumerate}
\item $I_{\rm pre}(\vec{x})$ has confocal elliptical isophotes with
  the radial brightness profile $S(x)$;
\item and if we choose \mbox{$f(\rho)=S(\rho)$} as \glam template;
\item and if $E_{\rm post}$ has only one minimum (non-degenerate).
\end{enumerate}
To explain, due to 1. and 2. we find a vanishing residual
\begin{equation}
  \vec{R}_{\rm
    pre}=\vec{I}_{\rm pre}-A\vec{f}_\rho=0~~,~{\rm
    for}~\vec{p}=\vec{p}_{\rm pre}\,,
\end{equation}
at the minimum of $E_{\rm pre}$ and consequently \mbox{$E_{\rm
    pre}(\vec{p}_{\rm pre}|\vec{I}_{\rm pre})=0$}. At the same time
for any metric $\mat{U}$, we also have \mbox{$E_{\rm
    post}(\vec{p}_{\rm pre}|\vec{I}_{\rm post})=0$} because
$\vec{I}_{\rm post}-A\mat{L}\vec{f}_\rho=\mat{L}\vec{R}_{\rm pre}=0$
for \mbox{$\vec{p}=\vec{p}_{\rm pre}$}. Because of the lower bound
\mbox{$E_{\rm post}\ge0$}, these parameters $\vec{p}_{\rm pre}$ have
to coincide with a minimum of $E_{\rm post}$ and hence indeed
\mbox{$\vec{p}_{\rm post}=\vec{p}_{\rm pre}$}. Note that the previous
argument already holds for the weaker condition
\mbox{$\mat{L}\vec{R}_{\rm pre}=0$} so that a mismatch between
$\vec{I}_{\rm pre}$ and the template at $\vec{p}_{\rm pre}$ produces
no bias if the mapped residuals vanish in the post-seeeing frame. In
addition, if this is the only minimum of $E_{\rm post}$ then the
estimator $\vec{p}_{\rm post}$ is also uniquely defined (condition
3). An extreme example of a violation of condition 3 is the degenerate
case \mbox{$N_{\rm pix}=1$}: the observed image consists of only one
pixel. Then every parameter set $(\vec{x}_0,\epsilon,t)$ produces
\mbox{$E_{\rm post}=0$}, if $A$ is chosen correspondingly. But this
should be a rare case because it is not expected to occur for
\mbox{$N_{\rm pix}>6$}, thus for images that span over more pixels
than \glam parameters.

A realistic pre-seeing image is neither elliptical nor is our chosen
template $f(\rho)$ likely to perfectly fit the radial light profile of
the image, even if it were elliptical. This mismatch produces a bias
in $\vec{p}$ only if $\mat{L}$ is singular, and the magnitude of the
bias scales with the residual of the template fit in the pre-seeing
frame.  To see this, let $\tilde{\vec{I}}_{\rm pre}$ be the
best-fitting template $A\vec{f}_\rho$ with parameters $\vec{p}_{\rm
  pre}$. The residual of the template fit in the pre-seeing frame is
\mbox{$\vec{R}_{\rm pre}=\vec{I}_{\rm pre}-\tilde{\vec{I}}_{\rm
    pre}$}.  In the vicinity of $\vec{p}_{\rm pre}$, we express the
linear change of $A\vec{f}_\rho$ for small $\delta\vec{p}$ by its
gradient at $\vec{p}_{\rm pre}$,
\begin{equation}
  \mat{G}=(\vec{G}_1,\ldots,\vec{G}_6)=\nabla_{\vec{p}}(A\vec{f}_\rho)\;,
\end{equation}
where
\begin{equation}
  \vec{G}_i=\left.\frac{\partial(A\vec{f}_\rho)}{\partial
      p_i}\right|_{\vec{p}=\vec{p}_{\rm pre}}
  ~,\,{\rm for}~i=1,\ldots,6\;.
\end{equation}
Each column $\vec{G}_i$ of the matrix $\mat{G}$ denotes the change of
$A\vec{f}_\rho$ with respect to $p_i$. Therefore, close to
$\vec{p}_{\rm pre}$ we find the Taylor expansion
\begin{eqnarray}
 \nonumber
  A\vec{f}_\rho&=&\tilde{\vec{I}}_{\rm pre}+\sum_{i=1}^6\vec{G}_i\,\delta
  p_i+O(\delta p_i\,\delta p_j)\\
  &=& \label{eq:taylor}
  \tilde{\vec{I}}_{\rm pre}+\mat{G}\,\delta\vec{p}+O(\delta p_i\,\delta p_j)\;.
\end{eqnarray}
Furthermore, since $\vec{p}_{\rm pre}$ is a local minimum of $E_{\rm
  pre}$, Eq. \Ref{eq:highresE}, we find at the minimum the necessary
condition
\begin{equation}
  \label{eq:epremin}
  \nabla_{\vec{p}}E_{\rm
    pre}(\vec{p}|\vec{I}_{\rm pre})=\vec{0}
  \Longrightarrow
  \mat{G}^{\rm T}\vec{R}_{\rm pre}=\vec{0}\;.
\end{equation}
This means that the residual $\vec{R}_{\rm res}$ is orthogonal to
every $\vec{G}_i$. Now let
\mbox{$\vec{R}=\mat{L}\vec{R}_{\rm pre}=\vec{I}_{\rm
    post}-\mat{L}\tilde{\vec{I}}_{\rm pre}$}
be the residual mapped to the post-seeing frame. Then
\Ref{eq:functional2} in the vicinity to $\vec{p}_{\rm pre}$ is
approximately
\begin{equation}
  \label{eq:Epost}
  E_{\rm post}(\vec{p}_{\rm pre}+\delta\vec{p}|\vec{I}_{\rm post})
  \approx
  (\vec{R}-\mat{L}\mat{G}\delta\vec{p})^{\rm T}\mat{U}
  (\vec{R}-\mat{L}\mat{G}\delta\vec{p})\;,
\end{equation}
which we obtain by plugging \Ref{eq:taylor} into \Ref{eq:functional2}.
This approximation is good if the bias is small, i.e., if the minimum
$\vec{p}_{\rm post}$ of $E_{\rm post}(\vec{p}|\vec{I}_{\rm post})$ is
close to $\vec{p}_{\rm pre}$. As shown in \citet{aitken1934least} in
the context of minimum-variance estimators, the first-order bias
\mbox{$\delta\vec{p}_{\rm min}=\vec{p}_{\rm post}-\vec{p}_{\rm pre}$}
that minimises \Ref{eq:Epost} is then given by
\begin{equation}
  \label{eq:bias}
   \delta\vec{p}_{\rm min}=
  \left(\mat{G}^{\rm T}\mat{L}^{\rm T}\mat{ULG}\right)^{-1}\mat{G}^{\rm T}\mat{L}^{\rm T}\mat{U}\,\vec{R}
  =
  \left(\mat{G}^{\rm T}\mat{U}_L\mat{G}\right)^{-1}\mat{G}^{\rm
    T}\mat{U}_L\,\vec{R}_{\rm pre}\;,
\end{equation}
with \mbox{$\mat{U}_L:=\mat{L}^{\rm T}\mat{U}\mat{L}$}.  This
reiterates that the bias $\delta\vec{p}_{\rm min}$ always vanishes
either for vanishing residuals \mbox{$\vec{R}_{\rm pre}=\vec{0}$}, or
if $\mat{L}^{-1}$ exists \emph{and} we choose
\mbox{$\mat{U}=(\mat{L}\mat{L}^{\rm T})^{-1}$} as metric. The latter
follows from Eq. \Ref{eq:bias} with $\mat{U}_L=\mat{1}$ and
Eq. \Ref{eq:epremin}.  Note that $\delta\vec{p}_{\rm min}$ does not
change if we multiply $\mat{U}$ by a scalar $\lambda\ne0$. Thus any
metric $\lambda\mat{U}$ generates as much bias as $\mat{U}$.

With regard to an optimal metric $\mat{U}$, we conclude from
Eq. \Ref{eq:bias} and \mbox{$\mat{G}^{\rm T}\vec{R}_{\rm pre}=0$} that
we can minimise the bias by the choice of $\mat{U}$ for which
$\mat{U}_L\approx\mat{1}$, or if $\|\mat{L}^{\rm
  T}\mat{U}\mat{L}-\mat{1}\|^2$ is minimal with respect to the
Frobenius norm $\|\mat{Q}\|^2=\tr{\mat{Q}\mat{Q}^{\rm T}}$.  This
optimised choice of $\mat{U}$ corresponds to the so-called
pseudo-inverse $(\mat{L}\mat{L}^{\rm T})^+$ of $\mat{L}\mat{L}^{\rm
  T}$ \citep{1992nrca.book.....P}. The pseudo-inverse is the normal
inverse of $\mat{L}\mat{L}^{\rm T}$ if the latter is regular.

A practical computation of $\mat{U}=(\mat{L}\mat{L}^{\rm T})^+$ could
be attained by choosing a set of orthonormal basis function
$\vec{b}_i$ in the pre-seeing frame, or approximately a finite number
$N_{\rm base}$ of basis functions that sufficiently describes images
in the pre-seeing frame. For every basis function, one then computes
the images $\mat{L}\vec{b}_i$ and the matrix
\begin{equation}
  \label{eq:LL}
  \mat{L}\mat{L}^{\rm T}=
  \mat{L}\left(\sum_{i=1}^\infty\vec{b}_i^{}\vec{b}_i^{\rm T}\right)\mat{L}^{\rm T}
  \approx
  \sum_{i=1}^{N_{\rm base}}
  \left(\mat{L}\vec{b}_i\right)\left(\mat{L}\vec{b}_i\right)^{\rm T}\;.
\end{equation}
The pseudo-inverse of this matrix is the metric in the post-seeing
frame. 

Specifically, for images that are only pixellated the matrix
$\mat{L}\mat{L}^{\rm T}$ is diagonal. To see this, consider
pixellations that map points $\vec{e}_i$ in the pre-seeing frame to a
single pixels $\vec{e}^\prime(\vec{e}_i)$ in the post-seeing frame, or
\mbox{$\mat{L}\vec{e}_i=\vec{e}^\prime(\vec{e}_i)$}; both $\vec{e}_i$
and $\vec{e}^\prime(\vec{e}_i)$ are unit vectors from the standard
bases in the two frames. According to \Ref{eq:LL}, the matrix
\mbox{$\mat{L}\mat{L}^{\rm
    T}=\sum_i\vec{e}^\prime(\vec{e}_i)[\vec{e}^\prime(\vec{e}_i)]^{\rm
    T}$} is then always diagonal, typically proportional to the unit
matrix, and easily inverted to obtain the optimal metric
$\mat{U}$. Unfortunately, this $\mat{U}$ only makes the linear-order
bias \Ref{eq:bias} vanish while we still can have a higher-order bias
because of the singular $\mat{L}$.

\subsection{Similarity between model-based and moment-based
  techniques}
\label{sect:connections}

Through the formalism in the previous sections it becomes evident that
there is no fundamental difference between model-based techniques and
those involving adaptive moments. Model-based techniques perform fits
of model profiles to the observed image, whereas moment-based
ellipticities with the adaptive weight $f^\prime(\rho)$ are
equivalently obtained from the image by fitting the ellipticial
profile $f(\rho)$ to the image. This requires, however, the existence
of a unique minimum of the functional $E(\vec{p}|I)$ which we assume
throughout the paper. We note that the similarity also extends to the
special case of unweighted moments in Eqs. \Ref{eq:unwqij} and
\Ref{eq:uwx0} which are in principle obtained by fitting
\mbox{$f(\rho)=\rho$} since \mbox{$f^\prime(\rho)=1$} in this case.

Nonetheless one crucial difference to a model-fitting technique is
that a fit of $f(\rho)$ does not assume a perfect match to
$I(\vec{x})$: the functional $E(\vec{p}|I)$ needs to have a minimum,
but the fit is allowed to have residuals $\vec{R}_{\rm pre}$, i.e.,
\mbox{$E(\vec{p}|I)\ne0$} at the minimum. As shown, for the estimator
based on \Ref{eq:functional2} this may cause bias, Eq. \Ref{eq:bias},
but only when analysing post-seeing images hence for
\mbox{$\mat{L}\ne\mat{1}$}. In practice, different methodologies to
estimate the pre-seeing moment-based ellipticity certainly use
different approaches. Our choice of a forward-fitting estimator
\Ref{eq:functional2} is very specific but is optimal in the sense that
it is always unbiased for \mbox{$\mat{L}\vec{R}_{\rm pre}=0$} or
regular $\mat{L}$. Yet other options are conceivable. For instance, we
could estimate moments in the post-seeing frame first and try to map
those to the pre-seeing frame (e.g., H03 or
\citealt{2011MNRAS.412.1552M} which aim at unweighted moments). It is
then unclear which weight is effectively applied in the pre-seeing
frame. Therefore the expression Eq. \Ref{eq:bias} for the bias is
strictly applicable only to our estimator and adaptive moments. But it
seems plausible that the bias of estimators of pre-seeing moments
generally depends on the residual $\vec{R}_{\rm pre}$ since the
brightness moments $x_{0,i}$ and $M_{ij}$ are the solutions to a
best-fit of elliptical templates.

In the literature the problem of bias due to residuals in model fits
is known as model bias or underfitting bias
\citep{2013MNRAS.434.1604Z,2010MNRAS.406.2793B}.  Consequently,
moment-based techniques are as prone to underfitting bias as
model-based methodologies.

\section{Statistical inference of ellipticity}
\label{sect:shearanalysis}

Realistic galaxy images $\vec{I}$ are superimposed by instrumental
noise $\delta\vec{I}$. Therefore the pre-seeing \glam ellipticity can
only be inferred statistically with uncertainties, and it is,
according to the foregoing discussion, subject to underfitting
bias. For a statistical model of the ellipticity $\epsilon$, we
exploit the previous conclusions according to which the ellipticity of
$I(\vec{x})$ for the adaptive weight $f^\prime(\rho)$ is equivalent to
$\epsilon$ of the best-fitting template $f(\rho)$. This renders the
inference of $\epsilon$ a standard forward-fit of a model
$A\mat{L}f(\rho)$ to $\vec{I}$.

We consider post-seeing images $\vec{I}$ with Gaussian noise
$\delta\vec{I}$, i.e., $\vec{I}=\vec{I}_{\rm post}+\delta\vec{I}$. The
covariance of the noise is
$\mat{N}=\ave{\delta\vec{I}\,\delta\vec{I}^{\rm T}}$, while
$\vec{I}_{\rm post}=\mat{L}\vec{I}_{\rm pre}$ is the noise-free image
in the post-seeing frame. A Gaussian noise model is a fair assumption
for faint galaxies in the sky-limited regime
\citep{2007MNRAS.382..315M}. Possible sources of noise are: read-out
noise, sky noise, photon noise, or faint objects that blend with the
galaxy image. If an approximate Gaussian model is not applicable, the
following model of the likelihood has to be modified accordingly.

The statistical model of noise are given by the likelihood ${\cal
  L}(\vec{I}|\vec{p})$ of an image $\vec{I}=\mat{L}\vec{I}_{\rm
  pre}+\delta\vec{I}$ given the \glam parameters $\vec{p}$.  We aim at
a Bayesian analysis for which we additionally quantify our prior
knowledge on parameters by the PDF $P_{\rm p}(\vec{p})$. We combine
likelihood and prior to produce the marginal posterior
\begin{equation}
  \label{eq:posterior}
  P_\epsilon(\epsilon|\vec{I})\propto
  \int\d A\,\d t\,\d^2x_0\;{\cal L}(\vec{I}|\vec{p})\,P_{\rm p}(\vec{p})
\end{equation}
of ellipticity by integrating out the nuisance parameters
$(\vec{x}_0,A,t)$; the constant normalisation of the posterior is
irrelevant for this paper but we assume that the posterior is proper
(it can be normalised). Our choice for the numerical experiments in
this study is a uniform prior $P_{\rm p}(\vec{p})$ for positive sizes
$t$ and amplitudes $A$, ellipticities \mbox{$|\epsilon|<1$}, and
centroid positions $\vec{x}_{0}$ inside the thumbnail image. As known
from previous Bayesian approaches to shear analyses, the choice of the
prior affects the consistency of the ellipticity posteriors (see,
e.g., BA14). The origin of the prior-dependence will become clear in
Sect. \ref{sect:glammarginal}.

With regard to notation, we occasionally have to draw random numbers
or vectors of random numbers $\vec{x}$ from a PDF $P(\vec{x})$ or a
conditional density $P(\vec{x}|\vec{y})$. We denote this by the
shorthand \mbox{$\vec{x}\sim P(\vec{x})$} and
\mbox{$\vec{x}\sim P(\vec{x}|\vec{y})$}, respectively. As
common in statistical notation, distinct conditional probability
functions may use the same symbol, as for instance the symbol $P$ in
$P(x|y)$ and $P(y|x)$.

\subsection{Caveat of point estimates}
\label{sect:toymodel}

\begin{figure}
  \begin{center} 
    \epsfig{file=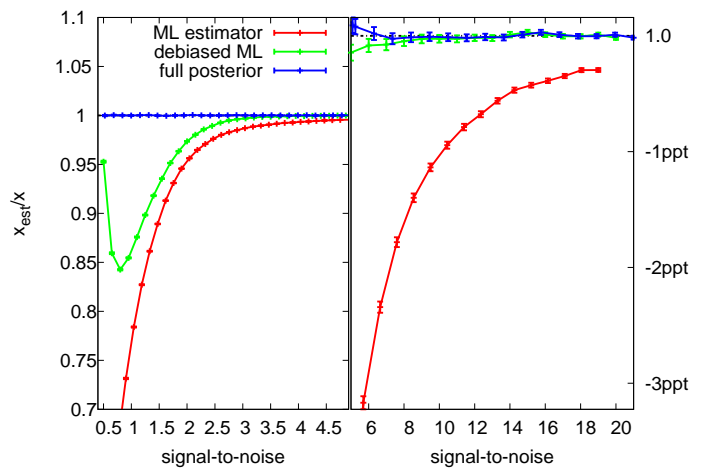,width=63mm,angle=-90}
  \end{center}
  \caption{\label{fig:toymodel} Toy-model demonstration of a
    maximum-likelihood estimator (red), a maximum-likelihood estimator
    with first-order bias correction (green), and an estimator
    exploiting the full posterior (blue). Data points display the
    estimator average ($y$-axis) over $10^6$ data points at varying
    signal-to-noise levels ($x$-axis). The true value to be estimated
    is \mbox{$x=1$}. The panels show different signal-to-noise
    regimes; $-n\rm ppt$ denotes $y=1-n/10^3$.}
\end{figure}

The bias in a lensing analysis is not only affected by how we
statistically infer galaxy shapes but also how we process the
statistical information later on. To demonstrate in this context the
disadvantage of point estimators in comparison to a fully Bayesian
treatment, we consider here a simplistic nonlinear toy model. This
model has one parameter $x$ and one single observable \mbox{$y=x^3+n$}
that is subject to noise $n$. By $n\sim N(0,\sigma)$
we draw random noise from a Gaussian distribution $N(0,\sigma)$
with mean zero and variance $\sigma$.  From the data $y$, we
statistically infer the original value of $x$. Towards this goal we
consider the (log-)likelihood of $y$ given $x$ which is $-2\ln{\cal
  L}(y|x)=(y-x^3)^2\,\sigma^{-2}+{\rm const}$.

A maximum likelihood estimator of $x$ is given by $x_{\rm
  est}=y^{1/3}$, the maximum of ${\cal L}(y|x)$. We determine the bias
of $x_{\rm est}$ as function of signal-to-noise ratio (S/N) $x/\sigma$
by averaging the estimates of \mbox{$N_{\rm real}=10^6$} independent
realisations of $y$. The averages and the standard errors are plotted
as red line in Fig. \ref{fig:toymodel}. Clearly, $x_{\rm est}$ is
increasingly biased low towards lower S/N levels.  In the context of
lensing, this would be noise bias. As an improvement we then correct
the bias by employing the first-order correction in
\citet{2012MNRAS.425.1951R} for each realisation of $y$. As seen in
the figure, this correction indeed reduces the systematic error, but
nevertheless breaks down for ${\rm S/N}\lesssim3$.

On the other hand in a fully Bayesian analysis, we obtain constraints
on $x$ that are consistent with the true value for any S/N. For this
purpose, we make $N_{\rm real}$ independent identically distributed
realisations (i.i.d.) $y_i$ and combine their posterior densities
\mbox{$P_{\rm post}(x|y_i)\propto{\cal L}(y_i|x)\,P_{\rm prior}(x)$}
by multiplying the likelihoods; we adopt a uniform (improper) prior
$P_{\rm prior}(x)=\rm const$. This gives us for
$\vec{y}=(y_1,\ldots,y_{N_{\rm real}})$, up to a normalisation
constant, the combined posterior
\begin{equation} 
  \ln{P_{\rm post}(x|\vec{y})}+{\rm const}= 
  \sum_{i=1}^{N_{\rm real}}\ln{{\cal L}(y_i|x)}
  =
  -\frac{1}{2\sigma^2}\sum_{i=1}^{N_{\rm
      real}}(y_i-x^3)^2\;.
\end{equation} 
As expected due to the asymptotic normality of posteriors (under
regularity conditions), for i.i.d. experiments $y_i$ the product
density is well approximated by a Gaussian $N(x_0,\sigma_x)$ and is
consistent with the true value $x$ \citep{vdV00}. We plot values of
$x_0$ and $\sigma_x$ in Fig. \ref{fig:toymodel} as blue data points.

In conclusion, keeping the full statistical information $P_{\rm
  post}(x|y)$ in the inference of $x$ yields consistent constraints
over the entire S/N range probed: the noise bias vanishes. Also note
that the Bayesian approach has not substantially increased the error
$\sigma_x$ compared to the error of the point estimator (relative
sizes of error bars); both approaches have similar efficiency.

\subsection{Likelihood model and underfitting bias}
\label{sect:glamnoise}

Inspired by the foregoing Bayesian toy model that is free of noise
bias, we set up a Bayesian approach for \glam ellipticities. To
construct a likelihood for a \glam fit in the pre-seeing frame, we
first consider, similar to Sect. \ref{sect:psfpixel}, the trivial case
where $\mat{L}$ is regular.  This is straightforward since we can map
the noisy image $\vec{I}\to\mat{L}^{-1}\vec{I}$ back to the pre-seeing
frame and determine the noise residual for given pre-seeing $\vec{p}$,
\begin{equation}
  \delta\vec{I}_{\rm pre}(\vec{p})=
  \mat{L}^{-1}\vec{I}-\vec{R}_{\rm pre}-A\vec{f}_\rho\;.
\end{equation}
The inverse noise covariance in the pre-seeing frame is
$\mat{L}^{\rm T}\mat{N}^{-1}\mat{L}$.  The logarithmic likelihood of
$\delta\vec{I}_{\rm pre}(\vec{p})$ in the Gaussian case is thus
\begin{eqnarray}
  \nonumber
  \lefteqn{-2\ln{{\cal L}_{\rm pre}(\vec{I}|\vec{p})+{\rm const}}}\\
  \nonumber
  &&=\delta\vec{I}_{\rm pre}(\vec{p})^{\rm T}\mat{L}^{\rm
    T}\,\mat{N}^{-1}\mat{L}\,\delta\vec{I}_{\rm pre}(\vec{p})
  \\
  \nonumber
  &&=\Big(\vec{I}-A\,\mat{L}\vec{f}_\rho-\mat{L}\vec{R}_{\rm
    pre}\Big)^{\rm T}\mat{N}^{-1}\,
  \Big(\vec{I}-A\,\mat{L}\vec{f}_\rho-\mat{L}\vec{R}_{\rm
    pre}\Big)
  \\
  \label{eq:likee0}
  &&=:\,\|\vec{I}-A\mat{L}\vec{f}_\rho-\mat{L}\vec{R}_{\rm
    pre}\|^2_{\mat{N}}\;,
\end{eqnarray}
where `const' expresses the normalisation of the
likelihood. Therefore, we can equivalently write the pre-seeing fit in
terms of available post-seeing quantities. Note here that the
transform $\mat{L}\vec{R}_{\rm pre}$ of the (unknown) pre-seeing
residual is for singular $\mat{L}$ and \mbox{$\vec{R}_{\rm pre}\ne0$}
not equal to the post-seeing residual $\vec{R}_{\rm post}$, which we
defined in a least-square fit of $A\mat{L}\vec{f}_\rho$ to
$\vec{I}_{\rm post}$ (see Sect. \ref{sect:psfpixel}).

In reality, $\mat{L}$ is singular so the previous steps cannot be
applied. Nevertheless, Eq. \Ref{eq:likee0} is the correctly specified
likelihood of a forward fit if the model
$\vec{m}(\vec{p}):=A\vec{f}_\rho+\vec{R}_{\rm pre}$ perfectly
describes the brightness profile $\vec{I}_{\rm pre}$ for some
parameters $\vec{p}_{\rm true}$. Therefore we expect no
inconsistencies for singular $\mat{L}$ as long as the correct
$\vec{R}_{\rm pre}$ can be given. Since $\vec{R}_{\rm pre}$ is
unknown, however, we investigate in the following the impact of a
misspecified likelihood that does not properly account for our
ignorance in the pre-seeing residuals. We do this by assuming
\mbox{$\vec{R}_{\rm pre}\equiv0$} and employing
\begin{equation}
  \label{eq:likee1}
  -2\ln{{\cal L}(\vec{I}|\vec{p})}+{\rm const}=
  \|\vec{I}-A\mat{L}\vec{f}_\rho\|^2_{\mat{N}}\;.
\end{equation}
Obviously, this is a reasonable approximation of \Ref{eq:likee0} if
\mbox{$\|\mat{L}\vec{R}_{\rm pre}\|\ll\|\vec{I}\|$} so that residuals
are only relevant if they are not small when compared to the image in
the post-seeing frame. It also follows from Eq. \Ref{eq:likee0} that
for \mbox{$\mat{L}\vec{R}_{\rm pre}=0$} the likelihood is correctly
specified even if \mbox{$\vec{R}_{\rm pre}\ne0$} and $\mat{L}$ being
singular, similar to the noise-free case. More generally we discuss
later in Sect. \ref{sect:discussion} how we could modify the
likelihood ${\cal L}(\vec{I}|\vec{p})$ to factor in our insufficient
knowledge about $\vec{R}_{\rm pre}$.  Until then the approximation
\Ref{eq:likee1} introduces underfitting bias into the likelihood
model, making it inconsistent with the true $\vec{p}_{\rm pre}$.

To show the inconsistency, we proceed as in the foregoing section on
the toy model. We consider a series of $n$ i.i.d. realisations
$\vec{I}_i$ of the same image $\vec{I}_{\rm post}$ and combine their
likelihoods ${\cal L}(\vec{I}_i|\vec{p})$ for a given $\vec{p}$ into
the joint (product) likelihood
\begin{equation}
  -2\ln{\prod_{i=1}^n{\cal L}(\vec{I}_i|\vec{p})}+
  {\rm const}:=
  \sum_{i=1}^n\|\vec{I}_i-A\mat{L}\vec{f}_\rho\|^2_{\mat{N}}\;,
\end{equation}
and we work out its limit for \mbox{$n\to\infty$}.  The joint
likelihood can be written as
\begin{eqnarray}
  \lefteqn{-2\ln{\prod_{i=1}^n{\cal L}(\vec{I}_i|\vec{p})}+{\rm const}}\\
  &=&\nonumber
  \sum_{i=1}^n\vec{I}_i^{\rm T}\mat{N}^{-1}\vec{I}_i-
  2\sum_{i=1}^n\vec{I}_i^{\rm T}\mat{N}^{-1}(A\mat{L}\vec{f}_\rho)
  +
  \sum_{i=1}^n(A\mat{L}\vec{f}_\rho)^{\rm T}\mat{N}^{-1}(A\mat{L}\vec{f}_\rho)\\
  &=&\nonumber
  \tr{\mat{N}^{-1}\,\sum_i\vec{I}_i^{}\vec{I}_i^{\rm T}}-
  2\sum_i\vec{I}_i^{\rm T}\,\mat{N}^{-1}(A\mat{L}\vec{f}_\rho)
  +
  n\,\|A\mat{L}\vec{f}_\rho\|^2_{\mat{N}}\;.
\end{eqnarray}
Here we have made use of the properties of the trace of matrices,
namely its linearity $\tr{\mat{A}+\mat{B}}=\tr{\mat{A}}+\tr{\mat{B}}$
and that $\vec{I}_i^{\rm T}\,\mat{A}\,\vec{I}_i=\tr{\vec{I}_i^{\rm
    T}\,\mat{A}\,\vec{I}_i}=\tr{\mat{A}\,\vec{I}_i\,\vec{I}_i^{\rm
    T}}$. For large $n$, we can employ the asymptotic expressions
\begin{equation}
  \label{eq:Niapprox}
  \sum_i\vec{I}_i^{\rm T}\to n\,\vec{I}_{\rm post}~;~
  \sum_i\vec{I}_i^{}\vec{I}_i^{\rm T}\to
  n\,\left(\mat{N}+\vec{I}_{\rm post}^{}\vec{I}_{\rm post}^{\rm T}\right)
\end{equation} 
to the following effect:
\begin{eqnarray}
  \label{eq:Ppbias}
  \lefteqn{-2\ln{\prod_{i=1}^n{\cal L}(\vec{I}_i|\vec{p})}+{\rm const}}\\
  &\to&\nonumber
  n\,\tr{\mat{N}^{-1}\mat{N}}+
  n\,\tr{\vec{I}^{}_{\rm post}\vec{I}_{\rm post}^{\rm T}\mat{N}^{-1}}\\
  &&\nonumber
  -2n\,\vec{I}_{\rm post}^{\rm T}\mat{N}^{-1}(A\mat{L}\vec{f}_\rho)
  +n\,(A\mat{L}\vec{f}_\rho)^{\rm T}\mat{N}^{-1}(A\mat{L}\vec{f}_\rho)\\
  &=&\nonumber
  n\,N_{\rm pix}+
  n\,\|\vec{I}_{\rm post}-A\mat{L}\vec{f}_\rho\|^2_{\mat{N}}\;.
\end{eqnarray}
Thus the joint likelihood peaks for \mbox{$n\to\infty$} at the minimum
$\vec{p}_{\rm post}$ of \mbox{$\|\vec{I}_{\rm
    post}-A\mat{L}\vec{f}_\rho\|_{\mat{N}}$}.  This is equivalent to
the location of the minimum of $E_{\rm post}(\vec{p}|\vec{I}_{\rm
  post})$, Eq. \Ref{eq:functional2}, with metric
$\mat{U}=\mat{N}^{-1}$. Consequently, as in the noise-free case, we
find an inconsistent likelihood if the residual $\vec{R}_{\rm pre}$ is
non-vanishing and if \mbox{$\mat{L}^{\rm T}\mat{N}^{-1}\mat{L}$} is
not proportional to the unity matrix. We additionally expect a smaller
bias $\delta\vec{p}=\vec{p}_{\rm post}-\vec{p}_{\rm pre}$ for smaller
levels of residuals.\footnote{Incidentally, the bias $\delta\vec{p}$
  in the likelihood and the underfitting bias in the noise-free case
  are equal for \mbox{$\mat{N}^{-1}\propto(\mat{L}\mat{L}^{\rm
      T})^+$}, which for regular $\mat{L}$ is equivalent to
  homogeneous, uncorrelated noise in the pre-seeing frame, which means
  we have \mbox{$\mat{L}^{\rm T}\mat{N}^{-1}\mat{L}\propto\mat{1}$}.}

With regards to the noise dependence of underfitting bias, we find
that $\delta\vec{p}$ does not change if we increase the overall level
of noise in the image $\vec{I}$. This can be seen by scaling the noise
covariance \mbox{$\mat{N}\mapsto\lambda\,\mat{N}$} with a scalar
$\lambda$. Any value \mbox{$\lambda>0$} results in the same minimum
location for $E_{\rm post}(\vec{p}|\vec{I}_{\rm post})$ so that
$\delta\vec{p}$ is independent of $\lambda$: there is no noise bias.

Moreover, the bias $\delta\vec{p}$ of a misspecified likelihood seems
to depend on our specific assumption of a Gaussian model for the
likelihood. It can be argued on the basis of general theorems on
consistency and asymptotic normality of posteriors, however, that for
\mbox{$n\to\infty$} we obtain the same results for other likelihood
models under certain regularity conditions (\citealt{vdV00}; Appendix
B in \citealt{gelman2003bayesian}). The latter requires continuous
likelihoods that are identifiable, hence \mbox{${\cal
    L}(\vec{I}|\vec{p}_1)\ne{\cal L}(\vec{I}|\vec{p}_2)$} for
\mbox{$\vec{p}_1\ne\vec{p}_2$}, and that the true $\vec{p}_{\rm true}$
is not at the boundary of the domain of all parameters $\vec{p}$.
This is stricter than our previous assumptions for the Gaussian model
where we needed only a unique global maximum of the likelihood ${\cal
  L}(\vec{I}_{\rm post}|\vec{p})$. It could therefore be that a unique
maximum is not sufficient for non-Gaussian likelihoods.

\subsection{Prior bias}
\label{sect:glammarginal}

The foregoing section discusses the consistency of the likelihood of a
\emph{single} image $\vec{I}$. We can interpret the analysis also in a
different way: if we actually had $n$ independent exposures
$\vec{I}_i$ of the same pre-seeing image, then combining the
information in all exposures results in a posterior \mbox{$P_{\rm
    p}(\vec{p}|\vec{I})\propto\prod_{i=1}^n\,{\cal
    L}(\vec{I}_i|\vec{p})\,P_{\rm p}(\vec{p})$} that is consistent
with $\vec{p}_{\rm post}$ for a uniform prior \mbox{$P_{\rm
    p}(\vec{p})=1$}. As discussed in \citet{vdV00}, the more general
Bernstein-von Mises theorem additionally shows that under regularity
conditions the choice of the prior is even irrelevant provided it sets
prior mass around $\vec{p}_{\rm post}$ (Cromwell's rule). This might
suggest that for a correctly specified likelihood ${\cal
  L}(\vec{I}_i|\vec{p})$, a fully Bayesian approach for the consistent
measurement of $\epsilon$ might be found that is independent of the
specifics of the prior and has no noise bias in the sense of
Sect. \ref{sect:glamnoise}. This is wrong as shown in the following.

Namely, in contrast to the previous simplistic scenario, sources in a
lensing survey have varying values of $\vec{p}$: they are
intrinsically different. For more realism, we therefore assume now
\mbox{$i=1\ldots n$} pre-seeing images that, on the one hand, shall
have different centroid positions $\vec{x}_{0,i}$, sizes $t_i$,
amplitudes $A_i$ but, on the other hand, have identical ellipticities
$\epsilon$. Our goal in this experiment is to infer $\epsilon$ from
independent image realisations \mbox{$\vec{I}_i=\vec{I}_{{\rm
      post},i}+\delta\vec{I}_i$} by marginalizing over the $4 n$
nuisance parameters $\vec{q}_i=(\vec{x}_{0,i},t_i,A_i)$. This
experiment is similar to the standard test for shear measurements
where a set of different pre-seeing images is considered whose
realisations $\vec{I}_i$ are subject to the same amount of shear
\citep[e.g.,][]{2010MNRAS.405.2044B}. As a matter of fact, the
inference of constant shear from an ensemble of images would just
result in $2n$ additional nuisance parameters for the intrinsic shapes
with essentially the same following calculations.

Let $P_{\rm q}(\vec{q}_i)$ be the prior for the four nuisance
parameters $\vec{q}_i$ of the $i$th image and $P_\epsilon(\epsilon)=1$
a uniform prior for $\epsilon$. We combine the \glam parameters in
$\vec{p}_i:=(\vec{q}_i,\epsilon)$, and we assume that all images have
the same prior density and that the noise covariance $\mat{N}$ applies
to all images. The marginal posterior of $\epsilon$ is then the
integral
  \begin{multline}
    \label{eq:Pproduct0}
    {\cal N}\,P_\epsilon(\epsilon|\vec{I}_1,\ldots,\vec{I}_n)=
    \prod_{i=1}^n\int\d^4q_i\;
    {\cal L}(\vec{I}_i|\vec{p}_i)\,P_{\rm q}(\vec{q}_i)\,P_\epsilon(\epsilon)\\
    =\int\d^4q_1\ldots\d^4q_n\;
    \prod_{i=1}^n{\cal L}(\vec{I}_i|\vec{p}_i)
    \times\prod_{i=1}^nP_{\rm q}(\vec{q}_i)\;,
  \end{multline}
  with ${\cal N}$ being a normalization constant.

The product of the likelihood densities inside the integral is given
by
\begin{eqnarray}
  \label{eq:Pproduct1}
  \lefteqn{-2\ln{\prod_{i=1}^n{\cal L}(\vec{I}_i|\vec{p}_i)+{\rm const}}}\\
  &&=\nonumber
  n\,\tr{\mat{N}^{-1}\frac{\sum_i\vec{I}_i^{}\vec{I}_i^{\rm
        T}}{n}}
  -2\,\tr{\mat{N}^{-1}\sum_i\vec{I}_i^{\rm
      T}(A_i\mat{L}\vec{f}_{\rho,i})}\\
  &&+\nonumber
  \sum_i\|A_i\mat{L}\vec{f}_{\rho,i}\|^2_{\mat{N}}\;.
\end{eqnarray}
Here we have taken into account that the \glam parameters partly
differ, indicated by the additional index in $A_i$ and
$\vec{f}_{\rho,i}$. This is different in Eq. \Ref{eq:Ppbias} where we
take the product of full likelihoods in $\vec{p}$-space without
marginalization. In the limit of $n\to\infty$, we find in addition
to the relations \Ref{eq:Niapprox} that
\begin{multline}
  \tr{\mat{N}^{-1}\sum_i\vec{I}_i^{\rm
      T}(A_i\mat{L}\vec{f}_{\rho,i})}=
  \tr{\mat{N}^{-1}\sum_i(\vec{I}_{{\rm post},i}+\delta\vec{I}_i)^{\rm
      T}(A_i\mat{L}\vec{f}_{\rho,i})}\\
  \to
  \tr{\mat{N}^{-1}\sum_i\vec{I}_{{\rm post},i}^{\rm T}\,A_i\mat{L}\vec{f}_{\rho,i}}
\end{multline}
because $\delta\vec{I}_i$ is uncorrelated to
$A\mat{L}\vec{f}_{\rho,i}$ so that $\delta\vec{I}_i^{\rm
  T}(A\mat{L}\vec{f}_{\rho,i})$ vanishes on average for many
$\delta\vec{I}_i$. Therefore, for the asymptotic statistic we can
replace all $\vec{I}_i$ by $\vec{I}_{{\rm post},i}$ in
Eq. \Ref{eq:Pproduct1} to obtain
\begin{equation}
  \prod_{i=1}^n{\cal L}(\vec{I}_i|\vec{p}_i)\to
  \prod_{i=1}^n{\cal L}(\vec{I}_{{\rm post},i}|\vec{p}_i)\;,
\end{equation}
and, as a result, for Eq. \Ref{eq:Pproduct0}
\begin{multline}
  \label{eq:limit}
  {\cal N}\,P_\epsilon(\epsilon|\vec{I}_1,\ldots,\vec{I}_n)\\
  \to
  \prod_{i=1}^n\int\d^4q_i\;
  {\cal L}(\vec{I}_{{\rm post},i}|\vec{p}_i)\,P_{\rm q}(\vec{q}_i)
  =:\prod_{i=1}^nP_\epsilon(\epsilon|\vec{I}_{{\rm post},i})\;,
\end{multline}
where $P_\epsilon(\epsilon|\vec{I}_{{\rm post},i})$ is the marginal
ellipticity posterior for the $i$th noise-free image.

The limit \Ref{eq:limit} has the interesting consequence that the
consistency with $\vec{p}_{\rm post}$ of the marginal posterior
depends on the specific choice of the prior density $P_{\rm
  q}(\vec{q}_i)$.  To show this, consider one particular case in
which, for simplicity, all pre-seeing images are identical such that
\mbox{$\vec{I}_{{\rm post},i}\equiv\vec{I}_{\rm post}$}.  Then,
according to \Ref{eq:limit}, the ellipticity posterior converges in
distribution to $[P_\epsilon(\epsilon|\vec{I}_{\rm post})]^n$ which
for \mbox{$n\to\infty$} peaks at the global maximum of
$P_\epsilon(\epsilon|\vec{I}_{\rm post})$.  It is then easy to see
that we can always change the position of this maximum by varying the
prior density in $P_\epsilon(\epsilon|\vec{I}_{\rm post})$. In
particular, even if the likelihoods are correctly specified, we
generally find an inconsistent marginal posterior depending on the
prior. A similar argument can be made if \mbox{$\vec{I}_{{\rm
      post},i}\ne \vec{I}_{{\rm post},j}$} for \mbox{$i\ne j$}.

\begin{figure}
  \begin{center}
    \epsfig{file=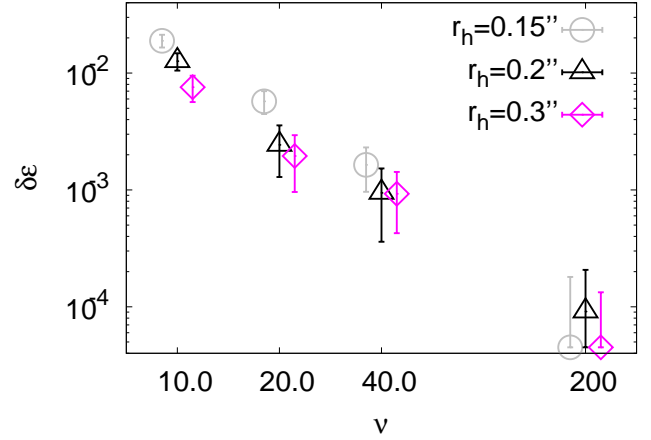,height=90mm,angle=-90}
  \end{center}
  \caption{\label{fig:biasbymarg} Prior bias in the marginal
      posterior $P_\epsilon(\epsilon|\vec{I}_1,\dots,\vec{I}_n)$ as
      function of S/N $\nu$ for different galaxy sizes $r_{\rm h}$ (in
      arcsec).  The posterior assumes a uniform prior. Shown is the
      error $\delta\epsilon=|\epsilon-\epsilon_{\rm true}|$ of the
      inferred $\epsilon$ for a true $\epsilon_{\rm true}=0.3$ as
      obtained by combining the marginal posteriors of $5\times10^3$
      exposures of the same galaxy with random centroid positions. The
      pixel size 0.1 arcsec equals the PSF size (Moffat). Galaxy
      profiles and \glam templates have a S\'ersic profile with
      $n=2$: there is no underfitting.}
\end{figure}

For a concrete example, we perform for Fig. \ref{fig:biasbymarg} a
simulated analysis of $5\times10^3$ noisy images with $\epsilon_{\rm
  true}=0.3$. All pre-seeing images are identical to one particular
template $A\vec{f}_\rho$ (S\'ersic profile with $n=2$). Therefore we
have a correctly specified likelihood model and no underfitting. We
adopt a uniform prior $P_{\rm q}(\vec{q})$ (and
$P_\epsilon(\epsilon)$). The details on the simulated images and their
analysis are given in Sect. \ref{sec:testsandresults}. For each data
point, we plot the mean and variance of the marginal posterior
$P_\epsilon(\epsilon|\vec{I}_1,\ldots,\vec{I}_n)$ relative to the true
ellipticity $\epsilon_{\rm true}$ as function of S/N $\nu$ and for
different image sizes $r_{\rm h}$. Evidently, the bias
$\delta\epsilon$ increases for smaller $\nu$ thereby producing a
noise-dependent bias which is not present when analysing the
consistency of ${\cal L}(\vec{I}|\vec{p})$ of individual images as in
Sect. \ref{sect:glamnoise}.

The sensitivity of the posterior to the prior implies that consistency
(for a correctly specified likelihood) could be regained by choosing
an appropriate prior density $P_{\rm q}(\vec{q}_i)$. On the other
hand, the dependence on the prior runs against the conventional wisdom
that the prior should become asymptotically irrelevant for
\mbox{$n\to\infty$} as the joint likelihood starts dominating the
information on inferred parameters. Indeed, general theorems show this
under certain regularity conditions \citep[see, e.g., discussion in
Chapter 4 of][]{gelman2003bayesian}. These conditions are, however,
not given here because the total number of model parameters is not
fixed but rather increases linearly with $n$ due to a new set of
nuisance parameters $\vec{q}_i$ for every new galaxy image
$\vec{I}_i$. The observed breakdown of consistency is the result. A
trivial (but practically useless) prior to fix the problem is one that
puts all prior mass at the true values of the nuisance parameters
which essentially leaves only $\epsilon$ as free model
parameter. Likewise, an analysis with sources that knowingly have the
same values for the nuisance parameters also yields consistent
constraints; this is exactly what is done in
Sect. \ref{sect:glamnoise}. A non-trivial solution, as reported by
BA14, is to use `correct priors' for the nuisance parameters that are
equal to the actual distribution of $\vec{q}$ in the sample. On the
downside, this raises the practical problem of obtaining correct
priors from observational data.

In summary, for an incorrect prior of $\vec{q}$, such as our uniform
prior, we find a noise-dependent bias in the inferred ellipticity
despite a fully Bayesian approach and a correctly specified
likelihood. We henceforth call this bias `prior bias'. This emphasizes
the difference to the noise bias found in the context of point
estimators.


\section{Simulated shear bias}
\label{sec:testsandresults}

We have classified two possible sources of bias in our Bayesian
implementation of adaptive moments: underfitting bias and prior
bias. In this section, we study their impact on the inference of
reduced shear $g$ by using samples of mock images with varying S/N,
galaxy sizes, and galaxy brightness profiles. In contrast to our
previous experiments, the images in each sample have random intrinsic
shapes but are all subject to the same amount of shear, and we perform
numerical experiments to quantify the bias. Moreover, this section
outlines practical details of a sampling technique for the posteriors
of \glam ellipticities and the reduced-shear, which is of interest for
future applications (Sect. \ref{sect:montecarloell} and
Sect. \ref{sect:gpost}).

The PSF shall be exactly known for these experiments. For a large
shear bias, we choose a relatively small PSF size and galaxy images
that are not much larger than a few pixels. According to the
discussion in Sect. \ref{sect:psfpixel}, underfitting bias is mainly a
result of pixellation which mathematically cannot be inverted.  Note
that a larger PSF size would reduce the underfitting bias since it
spreads out images over more image pixels. Overall the values for
shear bias presented here are larger than what is typically found in
realistic surveys \citep[e.g.,][]{2013MNRAS.434.1604Z}. We start this
section with a summary of our simulation specifications.

\subsection{Point-spread function}

If not stated otherwise, our simulated postage stamps consist of
$20\times20$ pixels (squares), of which one pixel has a size of
$0.1^\pprime$; the simulated galaxies are relatively small with a
half-light radius $r_{\rm h}$ of only a few times the pixel size
(\mbox{$0.15^\pprime\le r_{\rm h}\le0.3^\pprime$}). Additionally, we
adopt an isotropic Moffat PSF,
\begin{equation}
  I_{\rm psf}(\vec{x})\propto
  \left(1+\frac{|\vec{x}|^2}{\alpha^2}\right)^{-\beta}~~;~~
  \alpha:=\frac{\theta_{\rm
  FWHM}}{2\sqrt{2^{1/\beta}-1}}\;,
\end{equation}
with the full width half maximum (FWHM) of $\theta_{\rm
  FWHM}=0.1^\pprime$ and $\beta=5$ \citep{1969A&A.....3..455M}. 

\subsection{\glam template profiles}

As \glam templates we employ truncated S\'ersic-like profiles with
index $n$,
\begin{equation}
  \label{eq:sersic}
  f(\rho)=
  \exp{\left(-\left[\frac{\rho}{\rho_0}\right]^{\frac{1}{2n}}\right)}\,
  h(\!\sqrt{\rho})~;~
  h(x):=\frac{1}{\e^{5(x-3)}+1}
\end{equation}
and
\begin{equation}
  \rho_0:=(1.992n-0.3271)^{-2n}
\end{equation}
\citep{1989woga.conf..208C}. To avoid a numerical bias due to aliasing
at the edges of the grid during the Fourier transformation steps in
the sampling code, we have introduced the auxiliary function $h(x)$
that smoothly cuts off the S\'ersic profile beyond
\mbox{$\rho\approx9$}. If $f(\rho)$ were the radial light profile of
an elliptical galaxy, the truncation would be located at about three
half-light radii.

We use template profiles with \mbox{$n=2$} throughout. This index $n$
falls between the values of $n$ used for the model galaxies and thus
is a good compromise to minimise the fit residual and thereby the
underfitting bias.

\subsection{Mock galaxy images}
\label{sect:mocks}

We generate postage stamps of mock galaxy images with varying
half-light radii $r_{\rm h}$, radial light profiles, and
signal-to-noise ratios $\nu$. To this end, we utilise the code that
computes the post-seeing \glam templates (Appendix
\ref{sect:importancesampling}). We consider pre-seeing images of
galaxies with elliptical isophotes of three kinds of light profiles,
Eq. \Ref{eq:sersic}: (1) exponential profiles with S\'ersic index
\mbox{$n=1$} (EXP; exponential), (2) de-Vaucouleur profiles with
\mbox{$n=4$} (DEV; bulge-like) and (3) galaxies with profiles
\mbox{$n=2$} that match the profile of the \glam template (TMP;
template-like).  TMP galaxies hence cannot produce underfitting bias.

\begin{figure}
  \begin{center}
    \epsfig{file=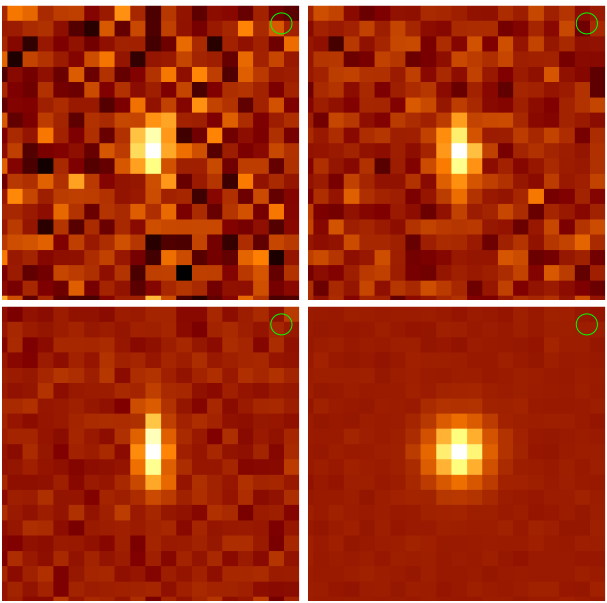,width=80mm,angle=0}
  \end{center}
  \caption{\label{fig:testexamples} Examples of simulated images of
    galaxies with random ellipticities. Signal-to-noise ratios from
    left to right and top to bottom: $\nu$=10, 20, 40, and 200.  The
    radial light profile of the pre-seeing images is exponential with
    $r_{\rm h}=0.2^\pprime$; the pixel size is $0.1^\pprime$. The FWHM
    of the PSF is indicated by the circle in the top right corners.}
\end{figure}

We devise uncorrelated Gaussian noise in the simulation of postage
stamps. To determine the RMS variance $\sigma_{\rm rms}$ of the pixel
noise for a given $\nu$, let $f_i$ be the flux inside image pixels
that is free of noise and \mbox{$f_{\rm tot}=\sum_i f_i$} the total
flux.  From this, we compute a half-light flux threshold $f_{\rm th}$
defined such that the integrated flux \mbox{$f_{\rm hl}=\sum_{f_i\ge
    f_{\rm th}}f_i$} above the threshold is \mbox{$f_{\rm hl}=f_{\rm
    tot}/2$} or as close as possible to this value. The pixels $i$
with \mbox{$f_i\ge f_{\rm th}$} are defined to be within the
half-light radius of the image; their number is \mbox{$N_{\rm
    hl}:=\sum_{f_i\ge f_{\rm th}}$}; the integrated noise variance
within the half-light radius is \mbox{$\sqrt{N_{\rm hl}}\,\sigma_{\rm
    rms}$}. The signal-to-noise ratio within the half-light radius is
therefore \mbox{$\nu=f_{\rm hl}^{}\,N_{\rm hl}^{-1/2}\,\sigma_{\rm
    rms}^{-1}$}, or
\begin{equation}
  \sigma_{\rm rms}=
  \frac{f_{\rm hl}}{\sqrt{N_{\rm hl}}\,\nu}\;.
\end{equation}
Figure \ref{fig:testexamples} depicts four examples with added noise
for different $\nu$.

To simulate galaxy images with intrinsic ellipticities that are
sheared by $g$, we make a random realisation of an intrinsic
ellipticity $\epsilon_{\rm s}$ and compute the pre-seeing ellipticity
with \Ref{eq:ss95}. As PDF for the intrinsic ellipticities we assume a
bivariate Gaussian with a variance of \mbox{$\sigma_\epsilon=0.3$} for
each ellipticity component; we truncate the PDF beyond
\mbox{$|\epsilon_{\rm s}|\ge1$}.

In realistic applications, centroid positions of galaxy images are
random within an image pixel, so we average all our following results
for bias over subpixel offsets within the quadratic solid angle of one
pixel at the centre of the image. This means: in an sample of mock
images, every image has a different subpixel position which is chosen
randomly from a uniform distribution. In this averaging process, care
has to be taken to perform an even sampling of subpixel offsets.  To
ensure this with a finite number of random points in all our following
tests, we employ a subrandom Sobol sequence in two dimensions for the
random centroid offsets \citep{1992nrca.book.....P}.

\subsection{Monte-Carlo sampling of ellipticity posterior}
\label{sect:montecarloell}

For practical applications of a Bayesian \glam analysis, we produce a
Monte Carlo sample of the posterior $P_\epsilon(\epsilon|\vec{I})$,
Eq. \Ref{eq:posterior} with the approximate likelihood ${\cal
  L}(\vec{I}|\vec{p})$ in Eq. \Ref{eq:likee1}. This sample consists of
a set $(\epsilon_i,w_i)$ of \mbox{$1\le i\le N_{\rm real}$} pairs of
values $\epsilon_i$ and $w_i$, which determine the sampling position
$\epsilon_i$ and a sampling weight $w_i$.  For this paper, we use
$N_{\rm real}=50$ sampling points. In contrast to a lensing analysis
with single-valued point estimators of ellipticity, a Bayesian
analysis employs the sample $(\epsilon_i,w_i)$ of each galaxy. To
attain the sample $(\epsilon_i,w_i)$ we invoke the importance sampling
technique \citep[e.g.,][]{Marshall56theuse,2010MNRAS.405.2381K}.

For this technique, we define an approximation $Q(\vec{p})$ of $P_{\rm
  p}(\vec{p}|\vec{I})$, the so-called importance distribution
function, from which we draw a random sample $\vec{p}_i$. The
ellipticity component $\epsilon_i$ of $\vec{p}_i$ is then assigned the
weight $w_i=P_{\rm p}(\vec{p}_i|\vec{I})\,Q(\vec{p}_i)^{-1}$ which we
normalise to $\sum_iw_i=1$ afterwards. As importance function, we use
a multivariate Gaussian with mean at the maximum $\vec{p}_{\rm ml}$ of
the likelihood ${\cal L}(\vec{I}|\vec{p})$ and a covariance defined by
the inverse Fisher matrix $\mat{F}^{-1}$ at the maximum
\citep{fisher1935,1997ApJ...480...22T}. More implementation details
are given in Appendix \ref{sect:importancesampling}. 

If $Q(\vec{p})$ is too different from $P_{\rm p}(\vec{p}|\vec{I})$,
the sample will be dominated by few points with large weights, usually
due to sampling points in the tail of the posterior for which the
importance function predicts a too low probability, i.e., $w_i$
becomes large. This is indicated by a small effective number $N_{\rm
  eff}=(\sum_iw_i^2)^{-1}$ of points compared to $N_{\rm real}$. This
can produce a poor convergence and thus extreme outliers in the
analysis that merely appear to have tight constrains on
$\epsilon$. Typically affected by this are images that are both small
compared to the pixel size and low in signal-to-noise. These images
tend to exhibit posteriors that allow parameter solutions with small
sizes $t$ compared to the true $t$ or high eccentricities
$|\epsilon|\gtrsim0.7$, giving the posterior a complex, distinctly
non-Gaussian shape. In future applications, this problem can be
addressed by finding a better model of the importance function.

For the scope of this paper, we address this problem at the expense of
computation time by increasing the number of sampling points and by
resampling. This means: for Monte-Carlo samples with \mbox{$N_{\rm
    eff}<N_{\rm real}/2$}, we draw more samples from the importance
distribution until \mbox{$N_{\rm eff}\ge N_{\rm real}/2$} in the
expanded sample. By an additional rule, we stop this process if the
expanded sample size becomes too large and reaches \mbox{$10\times
  N_{\rm real}$}. This case is indicative of a failure of the
importance sampling. If this happens, we switch to a time-consuming
standard Metropolis algorithm to sample the posterior with $10^3$
points after $10^2$ preliminary burn-in steps
\citep{1953JChPh..21.1087M}. This technique does not assume a
particular shape for the posterior but performs a random walk through
the $\vec{p}$ space, thereby producing a correlated sample along a
Monte-Carlo Markov chain. As symmetric proposal distribution for this
algorithm, we adopt a multivariate Gaussian with covariance
$1.5\times\mat{F}^{-1}$; all points of the chain have equal weight
$w_i$; the starting point of the chain is $\vec{p}_{\rm ml}$.
Finally, in the resampling phase, we bootstrap the expanded or
Metropolis sample by randomly drawing $N_{\rm real}$ points from it
with probability $w_i$ (with replacement). All selected data points
are given the equal weights \mbox{$w_i=N_{\rm real}^{-1}$} in the
final catalogue.

\subsection{Posterior density of reduced shear}
\label{sect:gpost}

For the inference of $g$, we convert the ellipticity posterior
$P_\epsilon(\epsilon|\vec{I})$ into a posterior $P_{\rm g}(g|\vec{I})$
of shear. To this end, let us first determine the $P_{\rm
  g}(g|\epsilon)$ for an exactly known $\epsilon$.  We express our
uncertainty on the intrinsic ellipticity $\epsilon_{\rm s}$ by the
prior $P_{\rm s}(\epsilon_{\rm s})$, and $P_{\rm g}(g)$ is our
a-priori information on $g$. The values of $g$ and $\epsilon_{\rm s}$
shall be statistically independent, i.e., $P_{\rm sg}(\epsilon_{\rm
  s},g)=P_{\rm s}(\epsilon_{\rm s})\,P_{\rm g}(g)$.\footnote{This
  assumption would be false if shear and intrinsic ellipticity were
  correlated, for instance due to selection effects related to the
  shape of a sheared image
  \citep{2003MNRAS.343..459H,2006MNRAS.368.1323H}. Thus correlations
  between intrinsic shapes and shear may affect Bayesian methodologies
  already at this level.}  Applying a marginalization over
$\epsilon_{\rm s}$ and then Bayes' rule, we find
\begin{multline}
  P_{\rm g}(g|\epsilon)\\
  =\int\d^2\epsilon_{\rm s}\;P_{\rm sg}(\epsilon_{\rm
    s},g|\epsilon) = \int\d^2\epsilon_{\rm s}\;
  \frac{P_{\epsilon}(\epsilon|g,\epsilon_{\rm s})\,P_{\rm
      sg}(\epsilon_{\rm s},g)} {{\cal N}(\epsilon)}\;,
\end{multline}
or, equivalently,
\begin{eqnarray}
  \label{eq:pgstep0}  
  {\cal N}(\epsilon)\,P_{\rm g}(g|\epsilon)&=&
  \int \d^2\epsilon_{\rm
    s}\;P_\epsilon(\epsilon|g,\epsilon_{\rm s})
  \,P_{\rm g}(g)\,P_{\rm s}(\epsilon_{\rm
    s})\\
  &=&\nonumber
  \int \d^2\epsilon_{\rm
    s}
  \;\delta_{\rm
    D}\Big(\epsilon-\epsilon(g,\epsilon_{\rm s})\Big)
  P_{\rm g}(g)\,P_{\rm s}(\epsilon_{\rm
    s})\\
  &=&\label{eq:pgstep1}  
  P_{\rm g}(g)
  \,P_{\rm s}\Big(\epsilon_{\rm s}(g,\epsilon)\Big)
  \left|\frac{\d^2\!\epsilon_{\rm s}}{\d^2\!\epsilon}\right|\;.
\end{eqnarray}
By ${\cal N}(\epsilon)$ we denote the normalisation of $P_{\rm
  g}(g|\epsilon)$,
\begin{equation}
  {\cal N}(\epsilon):=
  \int\d^2g\;P_{\rm g}(g)
  \,P_{\rm s}\Big(\epsilon_{\rm s}(g,\epsilon)\Big)
  \left|\frac{\d^2\!\epsilon_{\rm s}}{\d^2\!\epsilon}\right|\;.
\end{equation}
The normalisation ${\cal N}(\epsilon)$ only depends on the modulus of
$\epsilon$ for isotropic priors for symmetry reasons.  The integration
over the Dirac delta function $\delta_{\rm D}(x)$ uses
Eq. \Ref{eq:ss95}.  The determinant \mbox{$|\d^2\!\epsilon_{\rm
    s}/\d^2\!\epsilon|$}, the Jacobian of the linear conformal mapping
$\epsilon_{\rm s}(g,\epsilon)$, is a function of $\epsilon$ and $g$.
For the weak lensing regime \mbox{$|g|\le1$} of interest, this is
\begin{equation}
  \left|\frac{\d^2\!\epsilon_{\rm s}}{\d^2\!\epsilon}\right|=
  \frac{(1-|g|^2)^2}{|1-\epsilon\,g^\ast|^4}
\end{equation}
\citep{1998MNRAS.295..497G}.  To now account for the measurement
  error of $\epsilon$ in the shear posterior, we marginalize $P_{\rm
    g}(g|\epsilon)$, Eq. \Ref{eq:pgstep1}, over $\epsilon$ with
  $P_\epsilon(\epsilon|\vec{I})$ as error distribution,
\begin{eqnarray}
  \nonumber
  P_{\rm g}(g|\vec{I})&=&
  \int\d^2\epsilon\;
  P_{\rm
    g}(g|\epsilon)\,P_\epsilon(\epsilon|\vec{I})\\
  \label{eq:gtransform}
  &=&
  P_{\rm g}(g)\,(1-|g|^2)^2\,
  \int\d^2\epsilon\;
  \frac{P_{\rm s}\Big(\epsilon_{\rm s}(g,\epsilon)\Big)
    \,P_\epsilon(\epsilon|\vec{I})}
  {{\cal N}(\epsilon)\,|1-\epsilon\,g^\ast|^4}\;.
\end{eqnarray}

In our Monte Carlo scheme, we sample the ellipticity posterior of
every galaxy $i$ through $(\epsilon_{ij},w_{ij})\sim
P_\epsilon(\epsilon|\vec{I}_i)$ by $1\le j \le N_{\rm real}$ values
$\epsilon_{ij}$ of ellipticity and statistical weight $w_{ij}$. To
convert this sample to an approximation of the $g$ posterior, we
replace $P_\epsilon(\epsilon|\vec{I}_i)$ inside the integral
\Ref{eq:gtransform} by the point distribution
$\Sigma_j\,w_{ij}\,\delta_{\rm D}(\epsilon-\epsilon_{ij})$,
\begin{equation}
  \label{eq:gtransform2}
  P_{\rm g}(g|\vec{I}_i)
  \approx
  P_{\rm g}(g)\,(1-|g|^2)^2\,
  \sum_{j=1}^{N_{\rm real}}
    \frac{w_{ij}\,P_{\rm s}\!\left(\epsilon_{\rm s}(g,\epsilon_{ij})\right)}
  {{\cal N}(\epsilon_{ij})\,|1-\epsilon_{ij}g^\ast|^4}\;.
\end{equation}
For the following, we adopt a uniform prior for $g$, i.e., $P_{\rm
  g}(g)\propto H(1-|g|)$.  For reasons discussed in
Sect. \ref{sect:glammarginal}, the prior on the nuisance
$\epsilon_{\rm s}$ presumably has to be equal to the distribution of
intrinsic shapes for a consistent shear posterior. To investigate the
sensitivity of shear bias as to $P_{\rm s}(\epsilon_{\rm s})$ we
therefore use two types of priors: the correct prior, which is the
true intrinsic-shape distribution in Sect. \ref{sect:mocks}, or a
uniform prior $H(1-|\epsilon_{\rm s}|)$. For each prior $P_{\rm
  s}(\epsilon_{\rm s})$, we numerically compute ${\cal N}(\epsilon)$
for different $\epsilon$ once and interpolate between them later on in
\Ref{eq:gtransform2}.  We finally combine the posteriors $P_{\rm
  g}(g|\vec{I}_i)$ of all images $\vec{I}_i$ in the sample by
multiplying posterior values on a $g$-grid. For this, we apply
\Ref{eq:gtransform2} to every galaxy in the sample independently.

\subsection{Results}

\begin{figure*}
  \begin{center}
    \epsfig{file=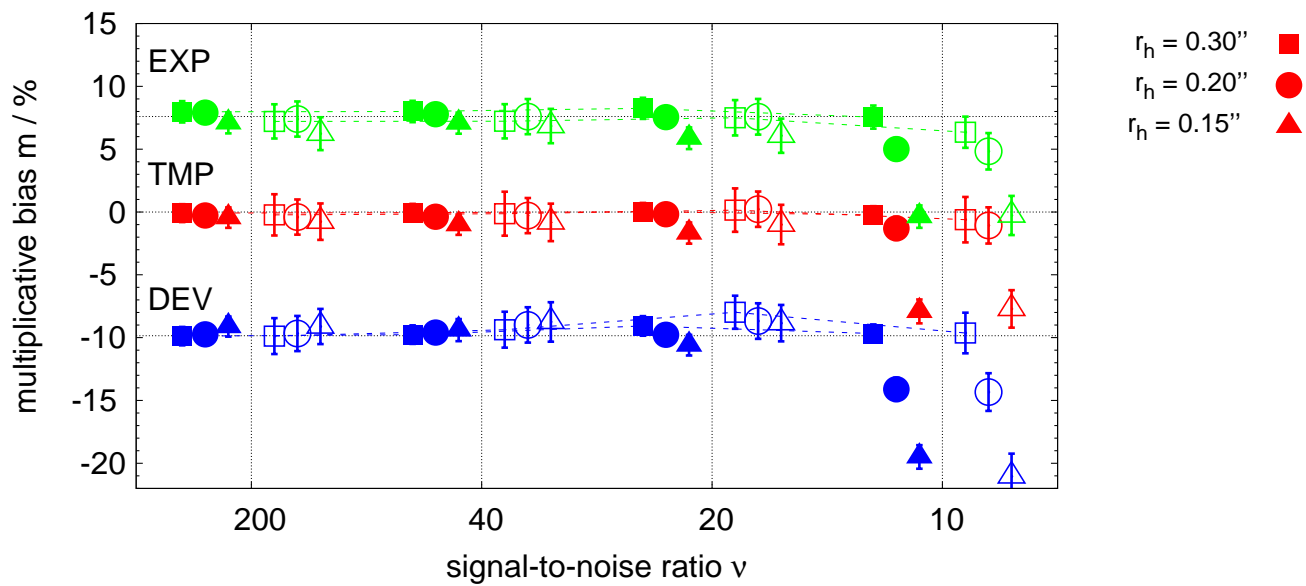,height=180mm,angle=-90}
  \end{center}
  \caption{\label{fig:glamtest2} Plots of the multiplicative bias $m$
    for simulated images, based on Eq. \Ref{eq:gtransform2}, for two
    different priors $P_{\rm s}(\epsilon_{\rm s})$ (filled data
    points: correct prior; open data points: uniform prior). Different
    styles for the data points indicate different image sizes $r_{\rm
      h}$, see key inside figure, while colors vary with galaxy type:
    \glam template (TMP; red); exponential (EXP; green); bulge-like
    (DEV; blue). Data points for $r_{\rm h}=0.3^\pprime$, same galaxy
    type, and same prior are connected by dotted lines to guide the
    eye. The prior for galaxy sizes $t$, amplitudes $A$, and centroids
    $\vec{x}_0$ is uniform giving rise to noise-dependent prior bias;
    the constant offset of EXP and DEV is due to underfitting. A
    square image pixel has the size $0.1^\pprime$, which also equals
    the PSF size. Results for $m$ for a larger PSF size can be found
    in Table \ref{tab:euclidlike}.}
\end{figure*}

For each experiment, we consider a sample of statistically independent
galaxy images $\hat{\vec{I}}=\{\vec{I}_1,\ldots,\vec{I}_n\}$ that are
subject to the same reduced shear $g$ but have random intrinsic
ellipticities $\epsilon_{\rm s}$. To infer $g$ from $\hat{\vec{I}}$ in
a fully Bayesian manner, we compute the posteriors $P_{\rm
  g}(g|\vec{I}_i)$ of $g$ for every image $\vec{I}_i$ separately with
Eq. \Ref{eq:gtransform2} and then combine all posteriors by the
product
\begin{equation}
  \label{eq:combinedPg}
  P_{\rm g}(g|\hat{\vec{I}})\propto
  \prod_{i=1}^nP_{\rm g}(g|\vec{I}_i)\;.
\end{equation}
This test is related, but not identical, to the ring test in
\citet{2007AJ....133.1763N}: we do not fix the ellipticity magnitude
$|\epsilon_{\rm s}|$ during one test run but instead use a random
sample of intrinsic ellipticities. Each galaxy sample consists of
\mbox{$n=5\times10^4$} simulated noisy galaxy images to typically
attain a precision for $g$ of the order \mbox{$10^{-3}$}. This number
applies for the non-uniform prior $P_{\rm s}(\epsilon_{\rm s})$. The
precision is lower for the less informative uniform prior.  Moreover,
for each galaxy with intrinsic $\epsilon_{\rm s}$ we also use
$-\epsilon_{\rm s}$ in another image belonging to the sample. By doing
so the mean of intrinsic ellipticities in an sample is always zero,
which substantially reduces the numerical noise in the result. For
every set of image parameters, we vary the input shear between the
three values \mbox{$g_{\rm true}\in\{0,\pm0.05,\pm0.1\}$}. For these
five measurements, let $\ave{g|g_{\rm true}}$ be the mean of the
posterior \Ref{eq:combinedPg} for a given $g_{\rm true}$ of the
sample. We then determine the multiplicative bias $m$ and the additive
bias $c$ in
\begin{equation}
  \label{eq:mcbias}
  \ave{g|g_{\rm true}}=
  (1+m)\,g_{\rm true}+c
\end{equation}
by best-fitting $m$ and $c$ to the mean and dispersion of
\Ref{eq:combinedPg} for the three input values of $g_{\rm true}$; $m$
shall be a real-valued number.

In Fig. \ref{fig:glamtest2}, we plot $m$ (filled data points)
juxtaposed with the corresponding values based on a uniform prior
(open data points); same data points indicate the same size $r_{\rm
  h}$ while colours, varying by row, indicate the light profiles of
the images. For the numbers in the table and the data points in the
figure, error bars quantify a 68\% posterior interval of the
measurement around the quoted mean. The size of error bars account for
pixel noise and shape noise. For the range of $\nu$ studied here, the
impact of pixel noise is subdominant: the errors increase only between
10\% and 20\% from $\nu=200$ to $\nu=10$; the increase is larger for
small images. Values for the additive bias $c$ (not shown) for all
profiles are small within a few $10^{-4}$, except at $\nu=10$ for DEV
with $r_{\rm h}=0.15^\pprime,0.2^\pprime$ and for TMP at
$0.15^\pprime$: there $c$ can reach negative amplitudes of the order
of $10^{-3}$. This indicates that an additive bias could be relevant
for image sizes close to the pixel size.

\begin{table}
  \caption{\label{tab:euclidlike} Multiplicative bias $m$ in per cent for three profiles TMP, EXP, DEV, and sizes $r_{\rm h}=0.2^\pprime$ only. The PSF size is $0.22^\pprime$ which is larger than in Fig. \ref{fig:glamtest2}. Values are quoted in per cent for different $\nu$ and for a correct prior $P_{\rm s}(\epsilon_{\rm s})$. The $1\sigma$ posterior errors are given only for TMP galaxies but are similar for the other galaxy models.}
  \begin{center}
    \begin{tabular}{cccc}
      \hline\hline
      $\nu$ & TMP & EXP & DEV \\
      \hline\\
      10 & $-22.1\pm5.0$ & $-12.8$ & $-17.8$\\
      20 & $-3.4\pm4.4$ & $+1.6$ & $-11.6$\\
      40 & $-1.7\pm4.2$ & $+5.3$ & $-6.8$\\
      200 & $+0.2\pm4.1$ & $+5.0$ & $-4.8$
    \end{tabular}
  \end{center}
\end{table}

In contrast, a multiplicative bias $m$ is typically significant. At
high S/N $\nu=200$, this bias in Fig. \ref{fig:glamtest2} is mainly
underfitting bias which is consequently absent for TMP galaxies since
their likelihood is correctly specified. The shear of bulge-like
galaxies DEV, with steeper slope compared to $f(\rho)$, is
underestimated by \mbox{$m=-9.6\pm0.5\,\%$}; the shear of exponential
galaxies EXP, which have a shallower slope, is overestimated by
\mbox{$m=+7.7\pm0.5\,\%$}. These numbers are the mean and standard
error for all galaxy sizes at \mbox{$\nu=200$}; they depend on the
specific PSF and pixel sizes chosen for this experiment.  

Compared to the underfitting bias at $\nu=200$, the bias $m$
significantly drops if \mbox{$\nu\lesssim20$} for all galaxies, most
prominently at $r_{\rm h}=0.15^\pprime$, but stays roughly constant
within a couple of per cent otherwise. We attribute the
noise-dependence of $m$ to prior bias owing to our choice of an
(incorrect) uniform prior for the nuisance parameters $\vec{q}$ of the
images (see Sect. \ref{sect:glammarginal}). The prior bias thus
becomes only relevant here at sufficiently low S/N. The particular
value of S/N is specific to the PSF size. This can be seen in Table
\ref{tab:euclidlike} which lists estimates for $m$ for galaxies, in a
slightly smaller sample \mbox{$n=2\times10^4$}, with size
\mbox{$r_{\rm h}=0.2^\pprime$} but for a modified, roughly twice as
large PSF size $0.22^\pprime$. Now the noise-dependence of prior bias
is already visible below \mbox{$\nu\lesssim40$}. On the other hand,
the underfitting bias, to be read off at high S/N \mbox{$\nu=200$}, is
roughly halved since the images are spread out over more pixels.

The choice of a uniform prior for the intrinsic ellipticities,
distinctively different from the true distribution of $\epsilon_{\rm
  s}$ in the sample, has a negligible effect on shear bias here for
both PSF sizes. Values of $m$ between the two priors $P_{\rm
  s}(\epsilon_{\rm s})$ are consistent, as found by comparing filled
and open data points of same colour and at same $\nu$ in
Fig. \ref{fig:glamtest2}.  Nevertheless, the resulting statistical
errors of shear, as opposed to its bias, are larger by roughly the
factor 1.7 with the uniform prior, which reflects the larger a-priori
uncertainty on $\epsilon_{\rm s}$. This implies that a shear bias due
to an incorrect prior $P_{\rm s}(\epsilon_{\rm s})$ is negligible in
this experiment.


\section{Discussion}
\label{sect:discussion}

Unlike galaxy ellipticities defined in terms of unweighted moments,
\glam provide a practical definition of ellipticity with useful
properties for any chosen adaptive weight $f^\prime(\rho)$: they (i)
are identical with the ellipticity of isophotes of elliptical galaxy
images; (ii) under the influence of shear they behave exactly like
$\epsilon$ defined with unweighted moments; (iii) they are unbiased
estimators of reduced shear. (i)-(iii) assume ideal conditions without
pixellation, PSF convolution, or pixel noise. These effects
fundamentally limit our ability to determine the \glam ellipticity
(see below). Under ideal conditions, see SS97, it is known that
$\epsilon$ for unweighted moments already obeys (iii). However,
unweighted moments are formally infinite for realistic galaxy images
because of pixel noise so that weighting is a necessity which is done
adaptively for \glam\!\!. For adaptive weights, we have shown (i) and
(ii) in Sect. \ref{sect:interpretation} and Sect. \ref{sect:shear}.
Their relevance as unbiased estimators for reduced shear, statement
(iii), follows thereafter from (ii) and the conclusions in SS97 for
unweighted moments. We emphasise that \glam do not require
``deweighting'' in a lensing analysis \citep{2011MNRAS.412.1552M}: the
ellipticity in term of the weighted moments is actually unbiased. In
particular, we do not have to devise the same weight function for all
galaxies as any weight function equally obeys (ii).  If the minimum of
the functional \Ref{eq:functional} uniquely exists, the moments
$(\vec{x}_0,\alpha\,\mat{M}_{\rm I})$ of the best-fitting template
$f(\rho)$ will, for a constant scalar $\alpha$, be the adaptively
weighted moments $(\vec{x}_0,\mat{M}_{\rm I})$ of the galaxy light
profile $I(\vec{x})$ for the weight $f^\prime(\rho)$ (Appendix
\ref{sect:gam}).  This means that the radial profile of the weight at
separation $r$ from the centroid is \mbox{$w(r)\propto r^{-1}\d
  f_r(r)/\d r$} for a template profile \mbox{$f_r(r):=f(r^2)$}. We
exploited this relation between adaptive moments and moments of a
best-fitting elliptical profile to analyse limits to adaptive
moment-based methods under the loss of image information.

Namely, for realistic images subject to pixellation and PSF effects,
underfitting bias can be present for \glam$\!$ if estimated from the
post-seeing image. It potentially arises because of a loss of
information on the pre-seeing galaxy image, fundamentally due to
pixellation. To explore the limitations, we assume noise-free images
and express in Sect. \ref{sect:psfpixel} the mapping between a
pre-seeing and post-seeing image by the linear operation
$\mat{L}$. The pre-seeing adaptive moments are equivalent to a
least-square fit with residual $\vec{R}_{\rm pre}$ of an elliptical
profile $f(\rho)$ to the pre-seeing image $I(\vec{x})$. If estimated
from the post-seeing image whilst ignoring residuals, the inferred
ellipticity is biased for \mbox{$\mat{L}\vec{R}_{\rm pre}\ne0$}
\emph{and} a singular $\mat{L}$. The latter indicates a loss of
information on the pre-seeing image.  For noise-free images and the
estimator \Ref{eq:functional2} the bias is, to linear order, given by
Eq. \Ref{eq:bias}. In \Ref{eq:functional2} the bias is optimally
reduced through the metric \mbox{$\mat{U}=(\mat{L}\mat{L}^{\rm
    T})^+$}. With this metric the estimator is unbiased for regular
$\mat{L}$, hence in principle also for an invertible convolution with
an anisotropic PSF; invertible convolutions have kernels $K(\vec{x})$
with \mbox{$\tilde{K}(\vec{\ell})\ne0$} in the Fourier domain.
Practically, however, a singular mapping is always present due to
pixellation so that the quadratic estimator \Ref{eq:functional2} could
never fully remove underfitting bias.

The inference of \glam ellipticity from noisy images with a likelihood
that ignores fitting residuals produces underfitting bias; the
underfitting bias depends on the correlation of pixel noise and its
homogeneity over the image. To deal with pixel noise in images, we
have put \glam in a Bayesian setting in Sect. \ref{sect:shearanalysis}
by employing an approximate model \Ref{eq:likee1} for the likelihood
that ignores fit residuals and thus equals an imperfect model fit with
an elliptical profile. We use this to explore the impact of a
misspecified likelihood that, as a result, can be shown to be subject
to underfitting bias. The bias is revealed in the limit of combining
the likelihoods of \mbox{$n\to\infty$} independent exposures of the
\emph{same} image (see Sect. \ref{sect:glamnoise}). To lowest order in
$\vec{R}_{\rm pre}$, we find the bias to vanish if \mbox{$\mat{L}^{\rm
    T}\mat{N}^{-1}\mat{L}$} is proportional to the unit matrix. The
underfitting bias is unchanged for a rescaled noise covariance: there
hence is no noise bias from a misspecified likelihood. Additionally,
the magnitude of bias depends on the details of $\mat{N}$ and can also
emerge for \mbox{$\mat{L}=\mat{1}$} if $\mat{N}$ is not proportional
to the unit matrix, which is the case for heterogeneous or correlated
noise. Due to the close relation of \glam ellipticity to other
definitions of ellipticity we expect similar dependencies of bias on
the noise covariance there, which may impact future calibration
strategies in lensing applications because they apparently should not
ignore cases of correlated or heterogeneous noise. Note that several
steps in the reduction of lensing data can produce correlated pixel
noise.

The underfitting bias from the likelihood can presumably be addressed
at the expense of an increased statistical error, such as by means of
a follow-up high-resolution survey that gathers statistical
information on the residuals $\vec{R}_{\rm pre}$.  First of all, one
obvious way to reduce underfitting bias is to choose a template that
provides a good fit to galaxy images to reduce the residuals. The
optimal template has the profile \mbox{$f_r(r)\propto S_r(r)$} for a
pre-seeing galaxy profile $S_r(r)$; this also optimises the
signal-to-noise ratio of the ellipticity
\citep{2002AJ....123..583B}. Even so, realistic galaxies can never be
perfectly fit by an elliptical model; there are always fitting
residuals $\vec{R}_{\rm pre}$. To deal with these residuals, we can
imagine a more elaborate Bayesian scheme where $\vec{R}_{\rm pre}$ are
additional nuisance parameters that we marginalize over by using an
empirical distribution $P(\vec{R}_{\rm pre}|\vec{p})$ of $\vec{R}_{\rm
  pre}$ given $\vec{p}$ (prior), i.e.,
\begin{equation}
  {\cal L}(\vec{I}|\vec{p})= 
  \int\d\vec{R}_{\rm
    pre}\;{\cal L}(\vec{I}|\vec{p},\vec{R}_{\rm pre})\,P(\vec{R}_{\rm
    pre}|\vec{p})\;,
\end{equation}
where ${\cal L}(\vec{I}|\vec{p},\vec{R}_{\rm pre})$ is the likelihood
of a post-seeing image, correctly specified by the post-seeing model
\mbox{$\vec{m}(\vec{p})=\mat{L}(A\,\vec{f}_\rho+\vec{R}_{\rm pre})$}
and the noise distribution. The marginalization would then increase
the statistical uncertainty in the posterior
$P_\epsilon(\epsilon|\vec{I})$ in Eq. \Ref{eq:posterior}. It is
conceivable to measure the prior in a dedicated survey by assessing
the distribution of residuals in template fits to high resolution,
high S/N images. This approach would be similar to that of BA14 where
a (correct) prior on galaxy-specific descriptors is needed. In
comparison to BA14, it is noteworthy here that the residuals
$\vec{R}_{\rm pre}$ in the pre-seeing frame are those of sheared
images for which we infer $\epsilon$; no assumptions on the unsheared,
source-plane images have to be made for this prior. All this, however,
still has to be tested, and it is not clear how this can be properly
implemented and if this marginalization is not the source of another
bias.

A consistent likelihood of single images is not sufficient for a
lensing analysis as new inconsistencies for $\epsilon$ or $g$ can
arise in samples of intrinsically different sources through incorrect
priors for intrinsic source parameters, such as sizes or centroid
positions (prior bias). Take for example our TMP galaxies in
Sect. \ref{sec:testsandresults} which have a correctly specified
likelihood by definition; they are free of underfitting bias. For the
two experiments presented in Fig. \ref{fig:biasbymarg} and
Fig. \ref{fig:glamtest2}, the source sample consists of pre-seeing
images with an unknown distribution of intrinsic parameters.  The
first experiment considers samples with constant ellipticity, the
second samples with constant reduced shear. Despite the absent
underfitting, noise bias now emerges in both cases if we apply our
uniform prior for the nuisance parameters in the marginal posterior of
either ellipticity or reduced shear; see TMP data points at
$\nu\lesssim20$.  We argue in Sect. \ref{sect:glammarginal} that the
emerging noise bias is related to the specific choice of priors and
can only be avoided by a special family of correct priors related to
the maximum of the posterior of noise-free images. According to
previous work, correct priors are the distributions of nuisance
parameters of the sources in the sample which have to be defined
externally (e.g., BA14, \citealt{2013MNRAS.429.2858M},
\citealt{2008MNRAS.390..149K}).  A conceivable alternative to external
priors, potentially worthwhile investigating in future studies, might
be to specify the priors down to a family of distributions, so-called
hyperpriors, with a finite, preferably small number of hyperparameters
\citep{gelman2003bayesian}. A hierarchical approach would then jointly
determine the posterior density of $\epsilon$ or $g$ and the
hyperparameters from all $\vec{I}_i$ alone in a fully Bayesian
way. Marginalizing over the hyperparameters accounts for the
uncertainty in the priors. Note that the incorrect uniform prior for
$\epsilon_{\rm s}$ in Fig. \ref{fig:glamtest2} has no significant
impact on the shear bias (compare open to filled data
points). Clearly, the importance of a correct prior depends on the
type of nuisance parameter.

That prior bias in Fig. \ref{fig:glamtest2} or Table
\ref{tab:euclidlike} becomes relevant only at low S/N can be
understood qualitatively. For a sufficiently high S/N, the mass of an
(identifiable) likelihood ${\cal L}(\vec{I}_{\rm post}|\vec{p})$
concentrates around $\vec{p}_{\rm post}$ such that the prior $P_{\rm
  p}(\vec{p})$ details have no strong impact on the marginal posterior
of one source. We also expect for high S/N the likelihood to be well
approximated by a normal Gaussian with maximum $\vec{p}_{\rm post}$
such that the marginalization over $\vec{q}=(A,t,\vec{x}_0)$
approximately yields a Gaussian posterior
$P_\epsilon(\epsilon|\vec{I}_{\rm post})$ with maximum near
$\epsilon_{\rm post}$, which is the true pre-seeing ellipticity for a
correctly specified likelihood. Note that the marginalization over
$\vec{q}$ of a multivariate Gaussian with maximum at
\mbox{$\vec{p}_0=(\epsilon_0,\vec{q}_0)$} produces another Gaussian
density with maximum at $\epsilon_0$. In addition, should the
likelihood be misspecified, $\epsilon_{\rm post}$ will be prone to
underfitting bias which is clearly visible by the offsets in $m$ for
EXP and DEV in Fig. \ref{fig:glamtest2}. 

Finally, we point out that while marginal posteriors
$P_\epsilon(\epsilon|\vec{I}_i)$, obtained on an image-by-image basis,
are sufficient to infer a constant ellipticity (or shear) it is not
obvious how to incoorporate them into a fully Bayesian analysis of
sources with varying shear. For example, in a fully Bayesian inference
of ellipticity correlations
$\xi_\vartheta=\ave{\epsilon_i^{}\epsilon_j^\ast}$ for images $ij$ at
separation $\vartheta$, we would compute the likelihood ${\cal
  L}(\hat{\vec{I}}|\xi_\vartheta)$ for given values of
\mbox{$\xi_\vartheta$} and a range of $\vartheta$, which principally
requires the repeated shape measurements for the entire sample of
images $\hat{\vec{I}}$ for every new $\xi_\vartheta$. This poses a
practical, computational problem. Proposed solutions to this problem
leave the path of a fully Bayesian analysis and commonly use unbiased
point estimators of $\epsilon_i$ on a image-by-image basis as input
for a statistical analysis \citep[e.g.,][]{2013MNRAS.429.2858M}. This
technique has essentially been applied to all lensing analyses so far,
but requires a calibration of the ellipticity estimates, especially
for the noise bias. The recently proposed hierarchical Bayesian
technique by \citet{2016MNRAS.455.4452A}, applied to data in
\citet{2017MNRAS.466.3272A}, is closest to a fully Bayesian analysis
in the above sense but also uses as input unbiased estimators of
source ellipticities with a multivariate Gaussian probability density
for the statistical errors of the estimates. Yet, it is surely
conceivable to extend this technique by directly modelling the
likelihood of source images $\vec{I}_i$. To reverse-propagate our
$P_\epsilon(\epsilon|\vec{I}_i)$ to a posterior of $\xi_\vartheta$, we
might be tempted here to Monte-Carlo a posterior PDF of
$\xi_\vartheta$ by independently drawing ellipticities
\mbox{$\epsilon_{im}\sim P_\epsilon(\epsilon|\vec{I}_i)$} from the
marginal posteriors to produce realisations of a joint vector
$\vec{\epsilon}_m=(\epsilon_{1m},\ldots,\epsilon_{nm})$, which could
be used with an estimator of $\xi_\vartheta$. Unfortunately, this
falsely assumes a separable joint posterior
\mbox{$P_\epsilon(\vec{\epsilon}|\hat{\vec{I}})=
  P_\epsilon(\epsilon_1|\vec{I}_1)\times \ldots\times
  P_\epsilon(\epsilon_n|\vec{I}_n)$} with statistically independent
ellipticities, potentially biasing (low) constraints on
$\xi_\vartheta$. As alternative we speculate about the following
hierarchical Bayesian scheme. We assume a parametric model
$P_\epsilon(\vec{\epsilon}|\hat{\vec{I}},\vec{\theta})$ for the joint
posterior density of ellipticities that is determined by (i) the
measured marginal ellipticities $P_\epsilon(\epsilon|\vec{I}_i)$ and
(ii) a set of unknown hyperparameters
$\vec{\theta}=(\theta_1,\theta_2,\ldots)$ that quantify the
correlations between the source ellipticities. A convenient choice in
this respect seem to be copula models that are specified exactly that
way \citep[e.g., Sect. 2 in][]{2010MNRAS.406.1830T}. We further
express our ignorance about $\theta_i$ by a (sufficiently
uninformative) hyperprior $P_\theta(\vec{\theta})$. We then produce
realisations $\vec{\epsilon}_m$ by first randomly drawing
$\vec{\theta}_m\sim P_\theta(\vec{\theta})$ followed by
$\vec{\epsilon}_{\rm m}\sim
P_\epsilon(\vec{\epsilon}|\hat{\vec{I}},\vec{\theta}_m)$.  A sample
$\{\vec{\epsilon}_1,\vec{\epsilon}_2,\ldots\}$ of joint ellipticities
is then utilized to propagate uncertainties on source ellipticities
and their mutual correlations through the lensing analysis, based on
an estimator for $\xi_\vartheta$ given a joint sample
$\vec{\epsilon}$. Note that in the limit $\nu\to\infty$ the hyperprior
for $\vec{\theta}$ becomes irrelevant because all marginal posteriors
$P_\epsilon(\epsilon|\vec{I}_i)$ will be sharply peaked at the true
ellipticity (if correctly specified).

\section*{Acknowledgments}

The authors are particularly grateful to Gary Bernstein who read and
commented on an earlier manuscript of this paper. The comments lead to
a substantial improvement.  We would also like to thank Tim Schrabback
for helpful discussions and comments, and our colleague Reiko Nakajima
for organising a weekly seminar on shape measurements in our institute
that triggered the research presented in this paper.  We also
acknowledge the comments by the referee of this paper which helped to
clarify and revise certain points, in particular the section on the
prior bias. This work has been supported by the Deutsche
Forschungsgemeinschaft in the framework of the Collaborative Research
Center TR33 `The Dark Universe'. Patrick Simon also acknowledges
support from the German Federal Ministry for Economic Affairs and
Energy (BMWi) provided via DLR under project no. 50QE1103.

\bibliographystyle{aa}
\bibliography{SimonSchneider_2016-29591}

\appendix

\section{Least-square conditions}
\label{sect:gam}

In this section, we derive general characteristics of \glam parameters
$\vec{p}=(A,\vec{x}_0,\mat{M})$ at a local minimum of the error
functional $E(\vec{p}|I)$, Eq. \Ref{eq:functional}. For convenience,
we introduce the bi-linear scalar product
\begin{equation}
  \label{eq:biform}
  \ave{g(\vec{x}),h(\vec{x})}:=
  \frac{1}{2}\int\d^2x\;g(\vec{x})h(\vec{x})
\end{equation}
to write the functional as 
\begin{equation}
  \label{eq:chi2}
  E(\vec{p}|I)=
  \Ave{I(\vec{x})-Af(\rho),I(\vec{x})-Af(\rho)}\;.
\end{equation}
Notice that $\rho:=(\vec{x}-\vec{x}_0)^{\rm
  T}\mat{M}^{-1}(\vec{x}-\vec{x}_0)$ is a function of $\vec{p}$. The
scalar product has the properties
\begin{eqnarray}
  \nonumber
  \ave{g(\vec{x}),h(\vec{x})}&=&\ave{h(\vec{x}),g(\vec{x})}\;;
  \\
  \nonumber
  \ave{a(\vec{p})\,g(\vec{x})+b(\vec{p})\,f(\vec{x}),h(\vec{x})}&=&\\
  \nonumber
  &&
  \hspace{-1cm}
  a(\vec{p})\ave{g(\vec{x}),h(\vec{x})}+b(\vec{p})\ave{f(\vec{x}),h(\vec{x})}\;;
  \\
  \nonumber
  \ave{a(\vec{p})g(\vec{x}),h(\vec{x})}&=&\ave{g(\vec{x}),a(\vec{p})h(\vec{x})}\;;
\end{eqnarray}
of which we make heavily use in the following.

Starting from here, we find as necessary condition for the
least-square solution of any parameter $p_i$:
\begin{eqnarray}
  \lefteqn{\frac{\partial E(\vec{p}|I)}{\partial p_i}=0}\\
  &&\nonumber
  =\Ave{-\frac{\partial[Af(\rho)]}{\partial p_i},I(\vec{x})-Af(\rho)}+
  \Ave{I(\vec{x})-Af(\rho),-\frac{\partial[Af(\rho)]}{\partial p_i}}\\
  &&\nonumber
  =-2\Ave{\frac{\partial[Af(\rho)]}{\partial
      p_i},I(\vec{x})-Af(\rho)}\\
  &&\nonumber
  =-2\Ave{\frac{\partial[Af(\rho)]}{\partial p_i},I(\vec{x})}+
  2A\Ave{\frac{\partial[Af(\rho)]}{\partial p_i},f(\rho)}\;,
\end{eqnarray}
or simply
\begin{equation}
  \label{eq:mlcondition}
  \Ave{\frac{\partial[Af(\rho)]}{\partial p_i},A^{-1}I(\vec{x})}=
  \Ave{\frac{\partial[Af(\rho)]}{\partial p_i},f(\rho)}
\end{equation}
that is fulfilled for all $p_i$. 

\subsection{Template normalisation}

From \Ref{eq:mlcondition} follows for the normalisation $p_i=A$ at the
minimum of $E(\vec{p}|I)$ that
\begin{equation}
  \label{eq:lsa}
  A=
  \frac{\ave{f(\rho),I(\vec{x})}}
  {\ave{f(\rho),f(\rho)}}  \;.
\end{equation}

\subsection{Image centroid position}

For the least-square centroid position $x_{0,i}$ with \mbox{$i=1,2$}
at the minimum of $E(\vec{p}|I)$, we find $\partial[Af(\rho)]/\partial
x_{0,i}=Af^\prime(\rho)\,\partial\rho/\partial x_{0,i}$ using
$f^\prime(\rho):=\d f(\rho)/\d\rho$. This can be expanded further by
considering
\begin{eqnarray}
  \label{eq:rhox0}
  \frac{\partial\rho}{\partial x_{0,i}}&=&
  \frac{\partial}{\partial x_{0,i}}
  \sum_{k,l}
  (x_k-x_{0,k})[\mat{M}^{-1}]_{kl}(x_l-x_{0,l})\\
  \nonumber
  &=&
  -\sum_{k,l}
  \left(
    \delta^{\rm K}_{ki}[\mat{M}^{-1}]_{il}(x_l-x_{0,l})+
    \delta^{\rm K}_{li}[\mat{M}^{-1}]_{ki}(x_k-x_{0,k})
  \right)\\
  \nonumber
  &=&
  -2\sum_k[\mat{M}^{-1}]_{ki}(x_k-x_{0,k})
  =-2\left[\mat{M}^{-1}(\vec{x}-\vec{x}_0)\right]_i\;.
\end{eqnarray}
Here we have exploited the symmetry
$[\mat{M}^{-1}]_{ij}=[\mat{M}^{-1}]_{ji}$ and introduced the Kronecker
symbol $\delta^{\rm K}_{ij}$. Therefore we find for the $i$th
component of the centroid position
\begin{equation}
  \Ave{Af^\prime(\rho)[\mat{M}^{-1}(\vec{x}-\vec{x}_0)]_i,f(\rho)-A^{-1}I(\vec{x})}=0
\end{equation}
or, equally, for both centroid components combined
\begin{equation}
  A\mat{M}^{-1}\ave{f^\prime(\rho)(\vec{x}-\vec{x}_0),f(\rho)}=
  \mat{M}^{-1}\ave{f^\prime(\rho)(\vec{x}-\vec{x}_0),I(\vec{x})}\;.
\end{equation}
The scalar product on the left-hand-side has to vanish due to symmetry
reasons. Therefore using the right-hand-side, we find at the minimum
\begin{equation}
  \label{eq:lsx0}
  \vec{x}_0=
  \frac{\ave{f^\prime(\rho)\vec{x},I(\vec{x})}}
  {\ave{f^\prime(\rho),I(\vec{x})}}\;.
\end{equation}
This shows that the centroid $\vec{x}_0$ at the least-square point is
the first-order moment $\vec{x}$ of the image $I(\vec{x})$ weighted
with the kernel $f^\prime(\rho)$. The weight is thus the derivative
$f^\prime(\rho)$ not $f(\rho)$.

\subsection{Image second-order moments}
\label{ap:beta}

For the components of $\mat{M}$, we consider the least-square values
of the inverse \mbox{$W_{ij}=[\mat{M}^{-1}]_{ij}$}, hence values
$W_{ij}$ for which $\partial [Af(\rho)]/\partial
W_{ij}=Af^\prime(\rho)\,\partial\rho/\partial W_{ij}$ vanishes. The
function $\rho$ can be expressed as
\begin{equation}
  \rho=
  {\rm tr}\big((\vec{x}-\vec{x}_0)^{\rm T}\mat{W}
  (\vec{x}-\vec{x}_0)\big)=
  {\rm tr}\left(\mat{W}\mat{X}\right)\;,
\end{equation}
using $\mat{X}:=(\vec{x}-\vec{x}_0)(\vec{x}-\vec{x}_0)^{\rm
  T}=\mat{X}^{\rm T}$ and the relation ${\rm
  tr}{(\mat{A}\mat{B}\mat{C})}={\rm tr}(\mat{B}\mat{C}\mat{A})$ for
traces of matrices. From this follows that $\partial\rho/\partial
W_{ij}=[\mat{X}]_{ij}=X_{ij}$, or 
\begin{eqnarray}
  A\Ave{f^\prime(\rho)X_{ij},A^{-1}I(\vec{x})}&=&
  A\Ave{f^\prime(\rho)X_{ij},f(\rho)}\\
  \nonumber
  \Longleftrightarrow~
  \Ave{f^\prime(\rho)X_{ij},I(\vec{x})}&=&
  \Ave{f^\prime(\rho)X_{ij},Af(\rho)}
\end{eqnarray}
by employing Eq. \Ref{eq:mlcondition}.  This set of (four) equations
can compactly be written as
\begin{equation}
  \Ave{f^\prime(\rho)\mat{X},Af(\rho)}=
    \Ave{f^\prime(\rho)\mat{X},I(\vec{x})}
\end{equation}
or utilising Eq. \Ref{eq:lsa}:
\begin{equation}
  \label{eq:lsm1}
  \frac{\ave{f^\prime(\rho)\mat{X},f(\rho)}}
  {\ave{f(\rho),f(\rho)}}
  =
  \frac{\ave{f^\prime(\rho)\mat{X},I(\vec{x})}}
  {\ave{f(\rho),I(\vec{x})}}\;.
\end{equation}
This last relation shows that the best-fitting template $f(\rho)$ has,
up to a constant scalar $\alpha$, the same central second-order
moments $\mat{M}_{\rm I}$ as the image,
\begin{eqnarray}
  \label{eq:lsm2}
  \mat{M}
  &=&
  \frac{\ave{f^\prime(\rho)\mat{X},f(\rho)}}
  {\ave{f^\prime(\rho),f(\rho)}}\\
  &=&
  \frac{\ave{f(\rho),f(\rho)}}
  {\ave{f^\prime(\rho),f(\rho)}}
  \frac{\ave{f^\prime(\rho),I(\vec{x})}}
  {\ave{f(\rho),I(\vec{x})}}
  \frac{\ave{f^\prime(\rho)\mat{X},I(\vec{x})}}
  {\ave{f^\prime(\rho),I(\vec{x})}}\\
  &=:&\label{eq:lsm2b}
  \alpha
  \frac{\ave{f^\prime(\rho)\mat{X},I(\vec{x})}}
  {\ave{f^\prime(\rho),I(\vec{x})}}
  =\alpha\mat{M}_{\rm I}\;.
\end{eqnarray}
In particular, the best-fitting template $Af(\rho)$ has, for a given
profile $f(s)$, the same ellipticity $\epsilon$ as the image
$I(\vec{x})$. Moreover, we derive from \Ref{eq:lsm1} the relation
\begin{equation}
  \mat{M}=
  \frac{\ave{f(\rho),f(\rho)}}{\ave{f^\prime(\rho),f(\rho)}}
  \frac{\ave{f^\prime(\rho)\mat{X},I(\vec{x})}}
  {\ave{f(\rho),I(\vec{x})}}
  =:
  \beta\frac{\ave{f^\prime(\rho)\mat{X},I(\vec{x})}}
  {\ave{f(\rho),I(\vec{x})}}\;.
\end{equation}
The scalar prefactor equals \mbox{$\beta=-2$} for
\mbox{$f(\rho)=\e^{-\rho/2}$}. Note that we have on the
right-hand-side $f(\rho)$ in the denominator rather than
$f^\prime(\rho)$.

We evolve the expression of $\beta$ a bit further by changing the
variables $\Delta\vec{x}=\mat{M}^{1/2}\vec{y}$ and hence
$\d^2\!x=\sqrt{\det{\mat{M}}}\,\d^2\!y$ for $\rho=\Delta\vec{x}^{\rm
  T}\mat{M}^{-1}\!\Delta\vec{x}$ in the integrals of the denominator
and nominator of $\beta$,
\begin{equation}
  \beta=
  \frac{\int\d^2y\;f^2(\vec{y}^{\rm T}\vec{y})}
  {\int\d^2y\;f^\prime(\vec{y}^{\rm T}\!\vec{y})f(\vec{y}^{\rm
      T}\!\vec{y})}
  =
  \frac{\int_0^\infty\d y\;y\,f^2(y^2)}
  {\int_0^\infty\d y\;y\,f^\prime(y^2)f(y^2)}
\end{equation}
or after another change of variables $y=\sqrt{s}$ and $\d y=\d
s/(2\sqrt{s})$,
\begin{equation}
  \beta=
  \frac{\ave{f(\rho),f(\rho)}}{\ave{f^\prime(\rho),f(\rho)}}=
  \frac{\int_0^\infty\frac{\d s}{2}\;f^2(s)}
  {\int_0^\infty\d s\;\frac{\d}{4\d s}f^2(s)}
  =
  -2\,\frac{\int_0^\infty\d s\;f^2(s)}
  {f^2(0)}\;,
\end{equation}
where the last step assumes that the profile $f(s)$ vanishes for
\mbox{$s\to\infty$}.

\section{Importance sampling of posterior distribution}
\label{sect:importancesampling}

\begin{algorithm}
  \caption{\label{alg:monteell} Sampling of the ellipticity posterior
    $P_\epsilon(\epsilon|\vec{I})$ for a given post-seeing
    galaxy image $\vec{I}$. Note that \glam parameters are
    \mbox{$\vec{p}=(A,\vec{x}_0,t,\epsilon)$}; $A$:
    amplitude; $\vec{x}_0$: centroid; $t$: size; $\epsilon$:
    ellipticity; assumption for residuals: $\vec{R}_{\rm
      pre}\equiv0$.}
  \SetKwInOut{Input}{Input}\SetKwInOut{Output}{Output} 
  \Input{\BlankLine $\vec{I}$: pixellated, noisy galaxy image\;
    $\mat{N}$: noise covariance of image\; $A\mat{L}\vec{f}_\rho$:
    post-seeing \glam templates with parameters $\vec{p}$\; $P_{\rm
      p}(\vec{p})$: prior of \glam parameters\;}
  \BlankLine
  \Output{\BlankLine $1\le i\le N_{\rm real}$ realisations
    $(\epsilon_i,w_i)$ with weights $w_i$; $N_{\rm eff}$:
    effective number of sampling points\;}
  \BlankLine 
  \Begin{ %
    $\vec{r}(\vec{p}):=\vec{I}-A\mat{L}\vec{f}_\rho$\; 
    $-2\ln{{\cal L}(\vec{I}|\vec{p})}:=\vec{r}^{\rm
      T}(\vec{p})\mat{N}^{-1}\vec{r}(\vec{p})$\; 
    \BlankLine 
    $\vec{p}_{\rm ml}\leftarrow -2\ln{{\cal L}(\vec{I}|\vec{p}_{\rm
        ml})}={\rm min.}$ (ML point)\;
    $F_{ij}\leftarrow \left.-\Ave{\frac{\partial^2\ln{{\cal
            L}(\vec{I}|\vec{p})}}{\partial p_i\partial p_j}\right|{\vec{p}_{\rm
        ml}}}$
    (Fisher matrix; Eq. \ref{eq:fisher})\;    
    \BlankLine
    \For{$i\leftarrow1$ \KwTo $N_{\rm real}$} 
    { 
      \Repeat{$|\epsilon_i|<1$ {\rm and} $P_{\rm p}(\vec{p}_i)>0$} {
        $\forall\,{1\le j\le6}~x_{ij }\sim N(0,1)$\;
        $\vec{p}_i\leftarrow\vec{p}_{\rm
          ml}+\sqrt{\mat{F}}^{-1}\vec{x}_i$\;
        $\epsilon_i\leftarrow\vec{p}_i=(A_i,\vec{x}_{0,i},t_i,\epsilon_i)$;
      }
      $\Delta\vec{p}_i\leftarrow\vec{p}_i-\vec{p}_{\rm ml}$\;
      $Q_i\leftarrow\exp{(-\Delta\vec{p}_i^{\rm T}\mat{F}\Delta\vec{p}_i/2)}$\; 
      $\pi_i\leftarrow{\cal L}(\vec{I}|\vec{p}_i)P_{\rm p}(\vec{p}_i)$\;
      $w^\prime_i\leftarrow\frac{\pi_i}{Q_i}$\;
    } 
    $\forall i~w_i\leftarrow \frac{w^\prime_i}{\sum_{i=1}^{N_{\rm
          real}}w^\prime_i}$\;
    $N_{\rm eff}\leftarrow \frac{1}{\sum_{i=1}^{N_{\rm real}}w_i^2}$\;
  }
\end{algorithm}

An overview of our algorithm for sampling the posterior $P_{\rm
  p}(\vec{p}|\vec{I})$ is given by Algorithm \ref{alg:monteell}, more
details follow below.

For the importance sampling of the posterior \glam variables
$\vec{p}$, we initially construct an analytical approximation
$Q(\vec{p})$ of the (full) posterior
\mbox{$P_{\rm p}(\vec{p}|\vec{I}):={\cal L}(\vec{I}|\vec{p})\,P_{\rm
    p}(\vec{p})$}
in Eq. \Ref{eq:posterior}, the so-called importance function.  Our
importance function is a local Gaussian approximation of
${\cal L}(\vec{I}|\vec{p})$ that is truncated for
\mbox{$|\epsilon|\ge1$}; $Q(\vec{p})$ is hence defined by its mean and
its covariance.  As previously discussed, we ignore residuals and set
\mbox{$\vec{R}_{\rm pre}\equiv0$} in Eq. \Ref{eq:likee0} for the
following.

For the mean $\vec{p}_{\rm ml}$ of the Gaussian $Q(\vec{p})$, we
numerically determine the maximum-likelihood (ML) point of ${\cal
  L}(\vec{I}|\vec{p})$ by minimising \Ref{eq:likee0} with respect to
$\vec{p}$. This is the most time consuming step in our algorithm,
which is done in the following three steps.
\begin{enumerate}
\item We quickly guess a starting point $\vec{p}_0$ of the centroid
  position $\vec{x}_0$ and the galaxy size $t$ by measuring the
  unweighted first and second-order moments of the post-seeing postage
  stamp. The initial ellipticity $\epsilon$ is set to zero. Herein we
  ignore the PSF. Alternatively, one could use estimates from object
  detection software such as \texttt{SExtractor} if available
  \citep{1996A&AS..117..393B}.
\item For an improved initial guess of the size $t$ and the centroid
  position $\vec{x}_0$, starting from $\vec{p}_0$ we perform a series
  of one-dimensional (1D) searches wherein we vary one parameter $p_i$
  while all other parameters $p_j$ with \mbox{$i\ne j$} are fixed
  \citep[coordinate descent;][]{coorddesc}. For each 1D search, we
  compute $-2\ln{{\cal L}(\vec{I}|\vec{p})}$ at five points over the
  range of values of $p_i$ that are allowed according to the prior
  $P_{\rm p}(\vec{p})$. We spline interpolate between the points to
  estimate the global minimum $p_i^{\rm min}$. Then we update $p_i$ to
  the value $p_i^{\rm min}$ and move on to the 1D search of the next
  parameter.  After every update we replace $A$ by its most likely
  value
\begin{equation}
  A_{\rm
    ml}=
  \frac{\vec{I}^{\rm T}\mat{N}^{-1}\mat{L}\vec{f}_\rho}
  {[\mat{L}\vec{f}_\rho]^{\rm T}\mat{N}^{-1}\mat{L}\vec{f}_\rho}
\end{equation}
(Eq. \ref{eq:lsa}) given the current values of
\mbox{$(\vec{x}_0,\epsilon,t)$}. We carry out the searches first
for the size $t$ and then for the two components of the centroid
position $\vec{x}_0$.
\item Starting from $\vec{p}$ in step 2, we apply the
  Levenberg-Marquardt algorithm (LMA) to iteratively approach
  $\vec{p}_{\rm ml}$ \citep{doi:10.1137/0111030}. This algorithm
  devises a linear approximation of $\mat{L}\vec{f}_\rho$ at $\vec{p}$
  based on the partial derivatives of $\mat{L}\vec{f}_\rho$.  Our
  implementation of the LMA searches only in the directions of
  $(\vec{x}_0,\epsilon,t)$ and again updates $A$ to $A_{\rm ml}$ after
  each iteration as in step 2. We find a slow convergence for large
  ellipticities, i.e., \mbox{$|\epsilon|\gtrsim0.3$}. This is
  improved, however, by adding an extra 1D search for the size $t$,
  $\epsilon_1$ and $\epsilon_2$ after each LMA iteration with an
  interval width of \mbox{$\pm0.1t$} (10 spline points) and $\pm0.1$
  (5 points) around the latest values of $t$ and $\epsilon_i$,
  respectively. For \mbox{$|\epsilon_n|\ge0.6$}, we use 10 instead of
  5 spline points which increases the accuracy for very elliptical
  objects.  We repeat the LMA until iteration \mbox{$n>5$} for which
  \mbox{$|\epsilon_{n-1}-\epsilon_{n}|\le10^{-4}$}; we change this
  threshold to $2\times10^{-5}$ once \mbox{$|\epsilon_n|\ge0.6$}. The
  latter is needed because the change of $\epsilon_n$ at each
  iteration becomes small for high $\epsilon$.
\end{enumerate}

For the covariance of $Q(\vec{p})$, we then numerically compute the
elements $F_{ij}$ of the (expected) Fisher matrix $\mat{F}$ around the
ML point $\vec{p}_{\rm ml}$,
\begin{equation}
  \label{eq:fisher}
  F_{ij}=
  \left(
    \frac{\partial[A\mat{L}\vec{f}_\rho]}
    {\partial p_i}
  \right)^{\rm T}
  \times\mat{N}^{-1}
  \times
  \left(
    \frac{\partial[A\mat{L}\vec{f}_\rho]}
    {\partial p_j}
  \right)\;,
\end{equation}
by considering partial derivatives of the best-fitting \glam templates
\citep[e.g.,][]{1997ApJ...480...22T}. This gives us
\begin{equation}
  Q(\vec{p})\propto
  \exp{\left(
      -\frac{1}{2}
      \left[\vec{p}-\vec{p}_{\rm ml}\right]^{\rm T}
      \mat{F}
      \left[\vec{p}-\vec{p}_{\rm ml}\right]
    \right)}\,
  H\left(P_{\rm p}(\vec{p})\right)\;,
\end{equation}
where $H(x)$ denotes the Heaviside step function. To quickly construct
the Fisher matrix, we employ the partial derivatives of
$\mat{L}\vec{f}_\rho$ in the last iteration of the previous LMA; the
derivative with respect to $A$ is simply $\mat{L}\vec{f}_\rho$.

For the main loop of the importance sampling, we randomly draw $N_{\rm
  real}$ points $\vec{p}_i\sim Q(\vec{p})$. Each
realisation $\vec{p}_i$ receives an importance weight
\mbox{$w_i=P_{\rm p}(\vec{p}_i|\vec{I})\,Q(\vec{p}_i)^{-1}$} for which
we have to evaluate the posterior $P_{\rm p}(\vec{p}_i|\vec{I})$;
neither $P_{\rm p}(\vec{p}|\vec{I})$ nor $Q(\vec{p})$ need to be
normalised here. In the end, we normalise all weights to unity for
convenience, i.e., \mbox{$\sum_iw_i\equiv1$}.  Moreover, for later
usage, we keep only the ellipticities $\epsilon_i$ of the parameter
sets $\vec{p}_i$. This Monte Carlo marginalizes over the uncertainties
of all other parameters.

As another technical detail, the above algorithm has to repeatedly
produce pixellated and PSF convolved versions of \glam templates
$\mat{L}\vec{f}_\rho$ for a given set of parameters $\vec{p}$. To
implement a fast computation in this regard, we construct the template
images on a high resolution (high-res) $256\times256$ grid in Fourier
space. To reduce boundary effects, we make the angular extent of the
high-res grid about 20 per cent larger than the extent of the
post-seeing galaxy image. Let $\tilde{c}_{\rm psf}(\vec{\ell})$ be the
Fourier transform of the PSF and
\begin{equation}
\tilde{f}(\ell)=
2\pi\int_0^\infty\d s\;f(s)\,J_0(s\ell)
\end{equation}
the transform of the radial template profile $f(\rho)$. Then the
Fourier coefficients of the PSF-convolved template are
\begin{equation}
  \label{eq:fourier}
  \widetilde{f}_{\rm psf}(\vec{\ell})=
  (\det{\mat{V}})\,\e^{{\rm
      i}\vec{\ell}\cdot\vec{x}_0}\tilde{c}_{\rm
    psf}(\vec{\ell})\tilde{f}(\vec{\ell}^{\rm T}\mat{V}^2\vec{\ell})\;.
\end{equation}
After setting up the grid of values in Fourier space, we
back-transform to coordinate space by means of the FFTW package
\citep{FFTW05}. Note that the Fourier coefficient at
\mbox{$\vec{\ell}=0$} has to be set to $\widetilde{f}_{\rm psf}(0)=
\det{V}\,\tilde{c}_{\rm psf}(0)\,\tilde{f}(0)$ to avoid a constant
offset of the template. After the FFT, we apply pixellation to produce
the vector $\mat{L}\vec{f}_\rho$ of post-seeing pixel values by
averaging pixels of the high-res grid within the extent of the
post-seeing image. We choose the angular extent of the high-res grid
such that an integer number of high-res pixels is enclosed by one
post-seeing pixel. For example, for our $20\times20$ post-seeing
images with pixel size \mbox{$r_{\rm post}=0.103125\approx0.1$}
arcsec, we use a high-res grid with extent of $2.4$ arcsec along each
axis, or a high-res pixel size of \mbox{$r_{\rm
    hres}=2.4/256=0.009375$} arcsec (square). Hence exactly $(r_{\rm
  post}/r_{\rm hres})^2=121$ high-res pixel correspond to one
post-seeing pixel.

Furthermore, to quickly compute the partial derivatives of
$\partial(\mat{L}\vec{f}_\rho)/\partial p_i$, needed for the LMA and
the Fisher matrix, we carry out the foregoing procedure, but we insert
the derivatives $\partial\widetilde{f}_{\rm psf}(\vec{\ell})/\partial
p_i$ as Fourier coefficients instead. These can be computed
analytically from Eq. \Ref{eq:fourier}.

\end{document}